\documentclass[twocolumn,prb,showpacs,preprintnumbers,amsmath,amssymb,floatfix]{revtex4-2}
\usepackage{graphicx}

\begin{document}
	
	\title{Vortex glass transition and thermal creep in niobium films}
	
	\author{Sameh M. Altanany, I. Zajcewa, T. Zajarniuk, A. Szewczyk and Marta Z. Cieplak}
	
	\affiliation{Institute of Physics, Polish Academy of Sciences, 02 668 Warsaw, Poland}

	\date{\today}
	

	\date{\today}
	
	\begin{abstract}
		{The evolution of the vortex glass (VG) phase transition and vortex creep with decreasing film thickness is studied in ultrathin, polycrystalline niobium films, with thickness in the range 7.4 nm to 44 nm, using current-voltage characteristics measurements in perpendicular magnetic field. Standard methods, including scaling laws, allow to identify VG transition in the thickest film, while in thinner films creep produces large uncertainty in the putative VG transition temperature and scaling exponents. Using strong pinning theory we perform analysis of the creep, and extract the dependence of the activation energy for vortex pinning on temperature, magnetic field, and film thickness. This analysis provides more information on vortex dynamics than the standard evaluation of critical current density. The results reveal two distinct regimes of pinning, which we propose to identify with $\delta l$ or $\delta T_c$-types of pinning (due to spacial fluctuation of mean free path $l$ or spacial fluctuation of superconducting transition temperature $T_c$, respectively). In the thickest film $\delta l$ pinning is observed, but with the decrease of film thickness the second pinning regime appears, and becomes dominant in the thinnest film. We link these pinning regimes with the structural disorder due to grain boundaries, which produce charge carrier scattering in the thickest film, but with decreasing film thickness gradually evolve into amorphous inclusions, producing fluctuations in the $T_c$.}
	\end{abstract}
	
	\maketitle
	
	\section {Introduction}

Vortex matter in the mixed phase of type-II superconductors is one of the most fascinating research topics in condensed matter physics. On the one hand, it provides a ground for testing of new concepts and theories, such as various ordered phases and phase transitions between them, with the primary example of melting of ordered Abrikosov vortex lattice in clean superconducting (SC) systems, or melting of vortex glass (VG) in systems with quenched disorder \cite{Fisher1991}. On the other hand, understanding of the impact of vortex pinning by defects on vortex dynamics in the presence of flowing currents is extremely important for applications, because flowing currents and thermal fluctuations contribute to creep-type vortex motion, which destroys dissipation-free transport \cite{Eley2021}.

Both the VG phase and flux creep phenomenon have been intensively studied in the past, mostly in high temperature superconductors where thermal fluctuations play an important role. Evaluation of these subjects is also important in case of ultrathin conventional SC films, which are frequently employed in nanoscale devices. Of particular interest is the role of disorder, which is introduced into the films with decreasing thickness, leading eventually to granular (or fractal) superconductivity \cite{Sacepe2020,Kapitulnik2019,Carbillet2020,Zaytseva2020}. Some of the SC properties in thin films with enhanced inhomogeneity have been evaluated recently, for instance, the Berezinskii-Kosterlitz-Thouless (BKT) transition in the absence of magnetic field \cite{Benfatto2009,Venditti2019,Weitzel2023}, or creep effects in the presence of magnetic field \cite{Eley2018}. However, while some studies of the VG transition and critical currents have been reported for various conventional SC films \cite{Villegas2005,Sun2013,Zhang2018,Song2019,Roy2019,Maccari2023}, the evolution of these properties with decreasing thickness is rarely studied. Few exceptions include study of VG transition in 100 nm Nb film and Nb/Cu superlattices \cite{Villegas2005}, and 3D-to-2D dimensional crossover of creep in Nb films with different thickness, all above 100 nm thick, studied by magnetization measurements \cite{Eley2018}.

In this work we use the current-voltage characteristics (IVC) measurements to examine the evolution of vortex dynamics with decreasing thickness of ultrathin, polycrystalline niobium (Nb) films, with thickness $d$ in the range 7.4 nm to 44 nm. This set of films has been studied by us before, and growing disorder with decreasing thickness has been documented by structural studies, leading eventually to structural transition to purely amorphous structure below about 3.3 nm \cite{Zaytseva2014}. Our goal in the present study is to observe evolution of the VG and pinning properties with the enhancement of disorder on the polycrystalline side of this transition, before transformation into amorphous structure.

In general, depending on the strength of vortex-defect interactions, two different pinning schemes are considered, weak collective pinning, induced by many atomic-size, weak point defects, which collectively give rise to an average pinning force \cite{Blatter1994}, and strong pinning, induced by relatively sparse but large defects, of the size comparable to the coherence length $\xi$ \cite{Ovchinnikov1991,vanderBeek2002}. Since the $\xi$ exceeds atomic-size point defects in our highly disordered films, we expect strong pinning to dominate. Indeed, as we will show, the IVC data in our films follow so-called excess current characteristics \cite{Thomann2012,Thomann2017}, which is a feature commonly observed in materials with strong pinning scenario \cite{Strnad1964,Kim1965,Berghuis1993,Xiao2002,Pace2004,Sacepe2019}.

Another classification of pinning relies on the mechanism of vortex-defect interaction, which may originate from the spatial variations of the SC transition temperature, $T_c$ ($\delta T_c$ pinning), or from variation in the mean free path $l$ of carriers, caused by scattering near lattice defects ($\delta l$ pinning); these mechanisms lead to different behaviors of the critical current density, $J_c$, with the change of temperature, magnetic field, or film thickness \cite{Griessen1994,Willa2016}. Therefore, the usual method of distinguishing between these mechanisms is to evaluate the dependence of $J_c$ on magnetic field, temperature, and film thickness. In our study we go beyond these type of evaluations. Namely, we use recently developed strong pinning theory in the presence of thermal fluctuations \cite{Buchacek2018,Buchacek2019,Buchacek2019_2} in order to extract most important creep parameter, activation energy for vortex pinning, $U_c$. We show that the analysis of $U_c$ dependence on temperature, magnetic field, and film thickness provides more insights into vortex dynamics in comparison with the simple analysis of the $J_c$. Our results suggests that the vortex pinning in our films evolves with film thinning, from $\delta l$ in thickest film, to $\delta l$ and $\delta T_c$ in different regions of phase diagrams in thinner films.
	
	\section {Experimental details}

	  The films were grown on glass substrates by magnetron sputtering in the high-vacuum chamber at room temperature, with niobium sandwiched between silicon buffer layers to avoid Nb oxidation, as already described \cite{Zaytseva2014}. The thickness of the Nb layers has been determined by low-angle X-ray reflectivity measurements. The detail structural studies of the films, reported previously \cite{Zaytseva2014,Demchenko2017}, reveal that the structure of the films changes from polycrystalline for $d \gtrsim 4$ nm, to purely amorphous for $d \lesssim 3.3$ nm. However, increasing disorder is seen in polycrystalline films on decreasing thickness. This is best visualized by the $d$-dependence of full width at half maximum (FWHM) of (110) diffraction peak for polycrystalline films, shown in the inset to Fig.\ref{Rsq}. We observe gradual increase of the FWHM, from about 0.37 deg in 44 nm film, to about 1.5 deg at 7 nm, and this is followed by more abrupt increase for $d \lesssim 7$ nm. The polycrystalline films contain uniformly thin layer (of thickness $\approx$ 1.5 nm) of amorphous Nb at the Nb/Si interface, formed in the initial stage of the Nb deposition \cite{Zaytseva2014}. In addition, small admixture of Si ions (5 to 10 at.\%) into Nb layer closest to the interface has been observed by X-ray Photoelectron Spectroscopy in the thinnest films \cite{Demchenko2017}.

In the present study we focus on polycrystalline films, with $d$ between 7.4 nm and 44 nm. Both the thin amorphous layer at the interface, and the tiny admixture of Si ions, introduce disorder into polycrystalline films, which is likely to play increasingly important role when $d$ is decreasing. Most importantly, polycrystalline films consist of grains, which are created during the film growth, with the grain boundaries affecting various film properties. For example, normal-state resistance ($R_N$) is observed to depend on film thickness as $R_N \sim d^{-2}$ \cite{Zaytseva2014}, what is predicted by boundary scattering theories \cite{Fuchs1938,Sondheimer1950}, and has also been observed in other thin Nb films \cite{Mayadas1972,Park1986,Yoshii1995}. The gradual increase of FWHM with decreasing $d$ signals decreasing size of polycrystalline grains, most likely accompanied by increasing width of grain boundaries, and possible appearance of amorphous regions between polycrystalline grains.

For transport measurements, the films are cut into $5\times5$ mm$^2$ size and they are photolithographically patterned into "Hall bar" structure, with the current path 200 $\mu$m wide and 2 mm long. Electrical contacts are made across the current and voltage channels at the film surface with the aid of indium solder. The sheet resistance ($R_{sq}$) and $I$-$V$ characteristics measurements were performed by a standard four-probe method in a magnetic field applied perpendicular to the film plane using Quantum Design Physical Property Measurement System (PPMS), with AC Transport (ACT) option.
	
\section {Results and Discussion}

\begin{figure}
\centering
\includegraphics[width=8.5cm]{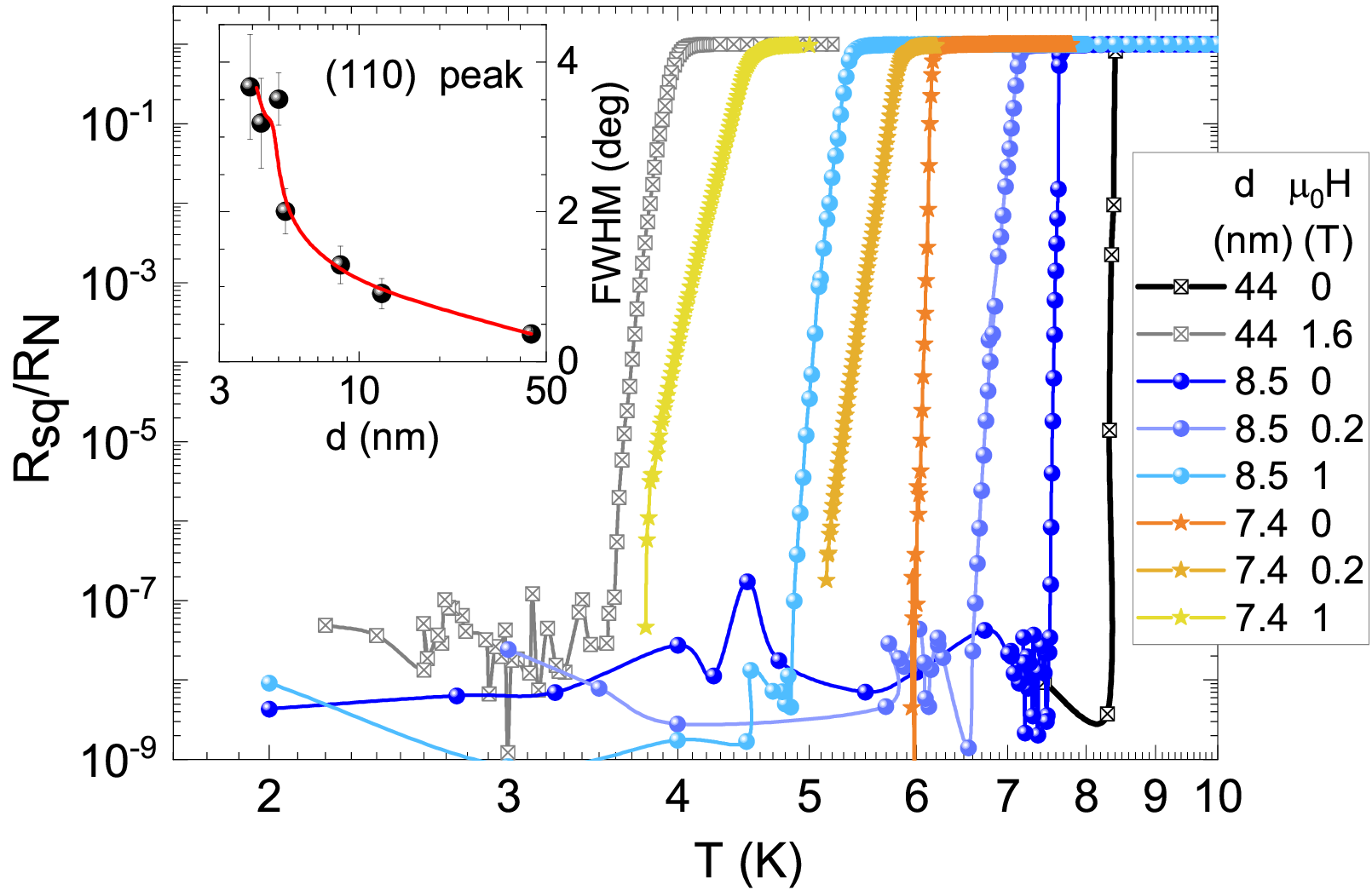}
\caption {Normalized sheet resistance $R_{sq}/R_N$ versus $T$ on a log-log scale, measured at $B$ ranging from 0 to 1.6 T for Nb films with $d$ = 7.4 nm, 8.5 nm, and 44 nm. Inset shows $d$-dependence of FWHM for (110) peak in X-ray diffraction spectrum.}\label{Rsq}
\end{figure}

\begin{table*}
 	\begin{tabular}{|c|c|c|c|c|c|c|c|c|c|c|}
 		\hline \begin{tabular}{c}
 			$d$ \\
 			(nm) \\
 		\end{tabular}
 		&\begin{tabular}{c}
 			$R_N$ \\
 			($\Omega$) \\
 		\end{tabular}
        &\begin{tabular}{c}
 			$T_c$ ($B$=0)\\
 			(K) \\
 		\end{tabular}
        &\begin{tabular}{c}
 			$\Delta T_c$ \\
 			(K) \\
 		\end{tabular}
        &\begin{tabular}{c}
 			$T_{BKT}$ \\
 			(K) \\
 		\end{tabular}
        &\begin{tabular}{c}
 			$ - dH_{c2}/dT|_{T_c}$ \\
 			(mT/K) \\
 		\end{tabular}
 		&\begin{tabular}{c}
 			$\xi (0)$ \\
 			(nm) \\
   		\end{tabular}
 		&\begin{tabular}{c}
 			$B$ \\
 			(T) \\
        \end{tabular}
        &\begin{tabular}{c}
 			$T_g$ \\
 			(K) \\
 		\end{tabular}
        &\begin{tabular}{c}
 			$z \pm 0.5$ \\
 		\end{tabular}
 		&\begin{tabular}{c}
 			$\nu \pm 0.7$ \\
 		\end{tabular}\\

  	    \hline 7.4 & 38.22 & 6.29 & 0.07  & 6.084 & 621 & 10.1 & 0.2  & 5.4  & 3.1   & 2.1 \\
 		           &       &      &       &       &     &      & 0.6  & 4.5  & 4.2   & 2.7 \\
  		           &       &      &       &       &     &      &  1   & 3.7  & 5.4   & 3 \\
                   &       &      &       &       &     &      &  1.5 & 2.9  & 5.4   & 2.7 \\
 	    \hline 8.5 & 10.69 & 7.71 & 0.059 & 7.625 & 442 & 10.8 & 0.1  & 7.15 & 3.75  & 2.2 \\
 	               &       &      &       &       &     &      & 0.2  & 6.82 & 4     & 3.1 \\
 		           &       &      &       &       &     &      &  1   & 5.0  & 5.5   & 2.2 \\
                   &       &      &       &       &     &      &  1.5 & 3.87 & 6     & 2.3 \\
                   &       &      &       &       &     &      &  2   & 2.65 & 4.9   & 2.7 \\
     	\hline 44  & 1.96  & 8.43 & 0.032 & 8.4   & 378 & 11.2 & 1    & 5.35 & 6     & 1.5 \\
                   &       &      &       &       &     &      & 1.6  & 3.48 & 6.3   & 1.7 \\
  		\hline
 	\end{tabular}

 	\caption{Parameters of Nb films: thickness $d$, sheet resistance $R_N$ at 10 K, $T_c$ (at $R_{sq}/R_N$=0.95) and width of the transition $\Delta T_c$ (both at $B=0$), $T_{BKT}$ determined from IVC at $B=0$, slope $dH_{c2}/dT|_{T_c}$ determined from linear fits to the upper critical field $H_{c2}$ in the vicinity of the $T_c$, coherence length $\xi (0)$, vortex glass transition temperature $T_g$ and critical exponents $z$ and $\nu$ from $I$-$V$ scaling analysis. The uncertainty of the $T_g$ value is about $\pm 0.25$ in the case of thinnest film, and about $\pm 0.2$ in case of remaining films.}
 	\label{table}
\end{table*}

Fig.\ref{Rsq} displays, on a double logarithmic scale, the temperature dependence of normalized sheet resistance $R_{sq}/R_N$ for three representative films, with $d$ = 7.4 nm, 8.5 nm, and 44 nm, in the presence of the magnetic field $B$ ranging between zero and 1.6 T. Here $R_N$ is the normal-state sheet resistance measured at $T = 10$ K. Table \ref{table} lists the $R_N$ and other parameters determined at $B=0$, the SC transition temperature $T_c$ at $R_{sq} /R_N = 0.95$, and $\Delta T_c$, the width of the transition defined by criterion 10\% to 90\% of $R_N$. It is observed that the $T_c$ decreases and the $\Delta T_c$ increases with decreasing $d$. Similar effect is observed when the magnetic field is increased, so that the region at which the resistance drops to zero is gradually shifted to lower $T$.

Using the value of the $T_c$ and the upper critical field, $H_{c2}$ (defined at $R_{sq}/R_N =0.95$), measured at small magnetic field perpendicular to the film surface in the vicinity of the $T_c$, we estimate the value of the $H_{c2} (0)$ and the in-plane coherence length $\xi (0) = {[\Phi_0 /2\pi H_{c2} (0)]}^{1/2}$ following pair-breaking theory \cite{Tinkham1985,Hsu1992}, $H_{c2} (0) = 0.825 \: dH_{c2}/dT|_{T_c} \: T_c$. Here $\Phi_0 =20.68$ Gs$\mu$m$^2$ is a flux quantum. The estimated values of the $\xi (0)$, shown in Table \ref{table}, agree well with the parameters of similar films studied by us previously \cite{Zaytseva2020}. Since the out-of-plane coherence length is likely to be smaller than $\xi (0)$, and bulk coherence length in Nb is 39.5 nm, we expect that films behave as two-dimensional (2D) or quasi-2D systems. Indeed, we have shown recently that the BKT transition is observed in some of these films \cite{Altanany2023}. $T_{BKT}$ listed in the table is defined as the temperature, at which IVC at $B=0$ follows the relation $V \thicksim I^n$ with $n=3$, as described in more detail in Ref. \cite{Altanany2023}.

In the next section in order to verify the possible VG nature of the zero-resistive state in these samples, we compare the resistance and the IVC experimental data with predictions for VG phase \cite{Fisher1989,Fisher1991}.

\subsection {Vortex glass transition and scaling laws}

We start our analysis with the resistance behavior. The usual method involves verification of the vanishing linear resistance ($R_{lin}$) on the approach to the $T_g$ from above \cite{Safar1992,Wagner1995}, as predicted by the VG theory in the very low current limit ($I \longrightarrow 0$) \cite{Koch1989,Fisher1991},

\begin{equation}
R_{lin}=(V/I)_{I\longrightarrow0} \propto(T-T_g)^{\nu(z+2-D)}
\label{Tg1}
\end{equation}

Here, $D$ is the dimensionality of the system, $\nu$ is the static exponent of the diverging vortex-glass correlation length $\xi_g\sim |T-T_g|^{-\nu}$, and $z$ is the dynamic exponent for the correlation time $\tau_g \sim \xi _g^z$. Accordingly, the established method to extract the $T_g$ and critical exponents \cite{Safar1992,Wagner1995} is to examine  logarithmic derivative of the sheet resistance which should follow linear $T$-dependence in the vicinity of the $T_g$,

\begin{equation}
	\left(\frac{d (\ln R_{sq})}{dT}\right)^{-1} = \frac{1}{\nu (z+2-D)}(T-T_g)
\label{Tg2}
\end{equation}

\begin{figure}
\centering
\includegraphics[width=6.5cm]{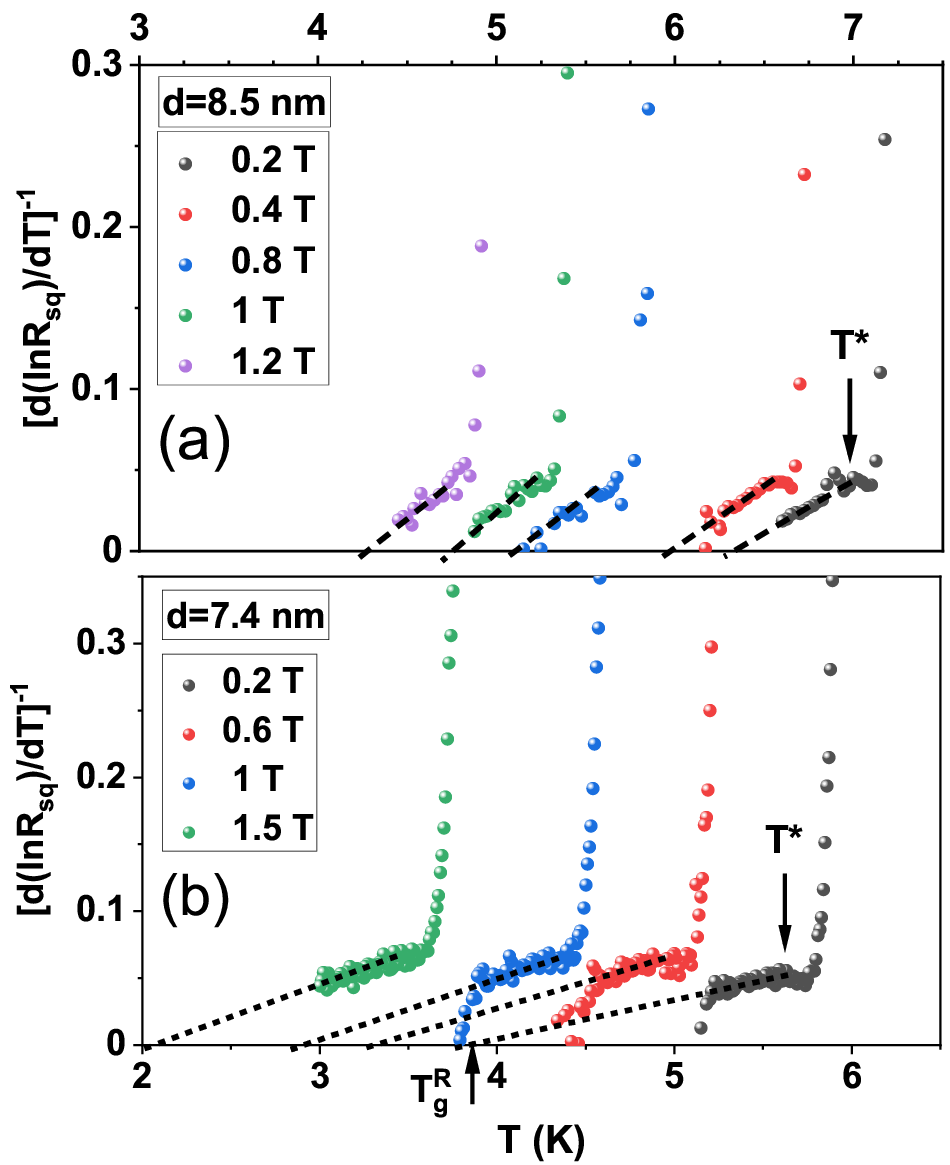}
\caption {(Color online) $[d (\ln R_{sq})/dT]^{-1}$ for different magnetic fields for films with $d=8.5$ nm (a) and $d=7.4$ nm (b). $T^{\ast}$ defines the upper bound of the critical VG region, and the dashed lines represent fits using Eq.(\ref{Tg2}) at $T < T^{\ast}$. The glass temperature, $T_g^R$, is determined from the cross of dashed line with the $T$ axis.}\label{dlnRsq}
\end{figure}

Figure \ref{dlnRsq} shows the plots of the quantity $[d (\ln R_{sq})/dT]^{-1}$ versus $T$ in the presence of applied magnetic fields for the two thinner films. The $R_{sq}(T)$ has been measured using low driving current in various magnetic fields, 0.5 mA in (a) and 0.2 mA in (b). We observe that all curves display linear portions below certain temperature $T^\ast$, which may be identified as the upper bound of the critical region associated with the VG thermodynamic transition \cite{Safar1992,Lee2010}. The fits to the linear portions are shown by dashed lines with the slopes, which gives the inverse of $\nu(z+2-D)$, reduced to $\nu z$ for $D=2$. In principle, from the cross of the dashed line with the $T$-axis one could determine the $T_g$ value, which we will call $T_g^R$, as marked in Fig.\ref{dlnRsq}. However, because the measuring current is finite, not zero, and because of disorder present in the samples, which may create distinct vortex behaviors in spatially limited regions of the films, the $T_g^R$ determined in this way may not really reflect true $T_g$ value as defined by Eq.(\ref{Tg1}). In order to get more insight into vortex dynamics, in the following we analyze the IVC's of the films.

\begin{figure}
\centering
\includegraphics[width=8cm]{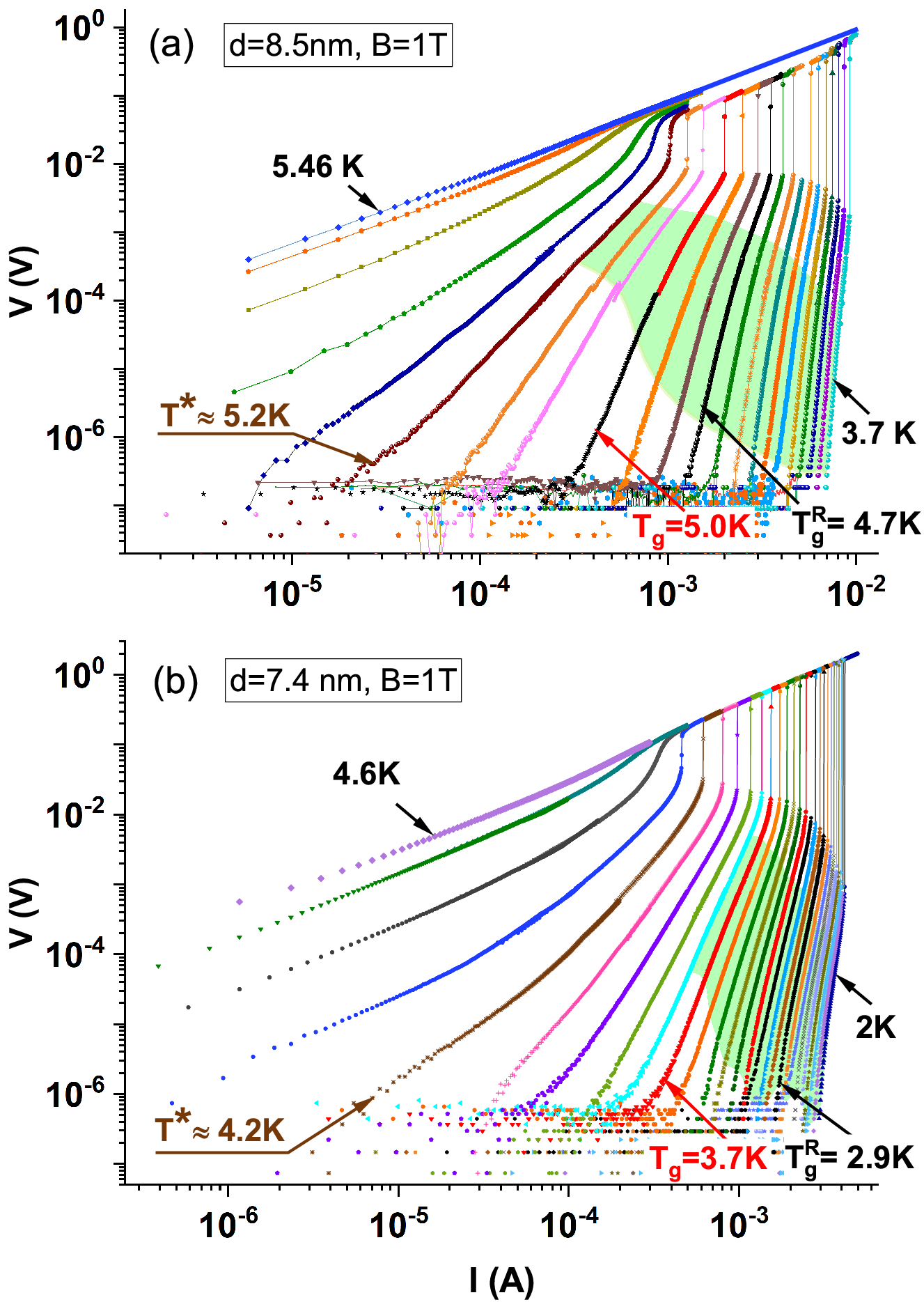}
\caption {IVC'c at various $T$ and $B = 1$ T for film with $d = 8.5$ nm, measured at $T$ from 3.7 K to 5.46 K (a), and for film with $d = 7.4$ nm, measured at $T$ from 2 K to 4.6 K (b). The black and red arrows show $T_g^R$'s and $T_g$'s determined by resistance and IVC's analysis, respectively. The light green areas in show the regions in which strong pinning theory \cite{Buchacek2019} describes satisfactorily the IVC's.}\label{IVC}
\end{figure}

Figure \ref{IVC} shows the IVC's on a log-log scale, measured for various temperatures at $\mu_0 H$ = 1 T in two thinner films. Arrows show the approximate position of $T^\ast$ and the location of $T_g^R$ isotherms as determined by resistance analysis. The isotherms measured at highest $T$ show linear dependence, $V \sim I$, in the whole range of current, indicating normal state. With $T$ decreasing down to $T^\ast$ the linear $V(I)$ dependence survives in the very low-current range, but it becomes nonlinear at higher current. With further decrease of $T$ below $T^\ast$ the linear dependence is no longer seen, instead the low-current $V(I)$ dependence changes into power-law, with the slope on the log-log graph increasing as $T_g^R$ is approached. We note, however, that the slope is not constant in the whole current range, but it becomes smaller at high current, suggesting the approach to the flux flow regime. This is followed by an abrupt jump towards the normal state.

These two high-current features, the jump and the change of slope, require some comments. As it is visible in Fig.\ref{IVC}(a) the jump in this film does not lead directly to a normal state, but to an intermediate resistive state, with resistance slightly below normal state resistance; such behavior has been reported previously for various systems, and we will come back later to discuss briefly the possible origin of this effect. The other feature, the change of slope, is best seen in areas marked by light green color; we will show in the next section that the data in these areas are affected by a flux creep, and they are well described by strong pinning theory, which predicts the shape of the IVC in the presence of strong pinning and thermal fluctuations \cite{Buchacek2018,Buchacek2019,Buchacek2019_2}.

Before turning to these issues, it is instructive to proceed first with the analysis of the IVC data using scaling laws of the VG theory, which are customarily done in order to validate the VG picture. The scaling law predicts that the $I$-$V$ curves near the $T_g$ can be scaled into two distinct branches ($f_\pm$) above and below the $T_g$,

\begin{equation}
\frac {V}{I(T-T_g)^{\nu(z+2-D)}} = f_\pm\left(\frac{I}{|T-T_g|^{\nu(D-1)}}\right),
\label{scaling}
\end{equation}

At the $T_g$ the $I$-$V$ isotherm should satisfy the power law:
	
\begin{equation}
V(I)\sim I^{(z+1)/(D-1)}.
\label{zvalue}
\end{equation}

\begin{figure*}
\centering
\includegraphics[width=18cm]{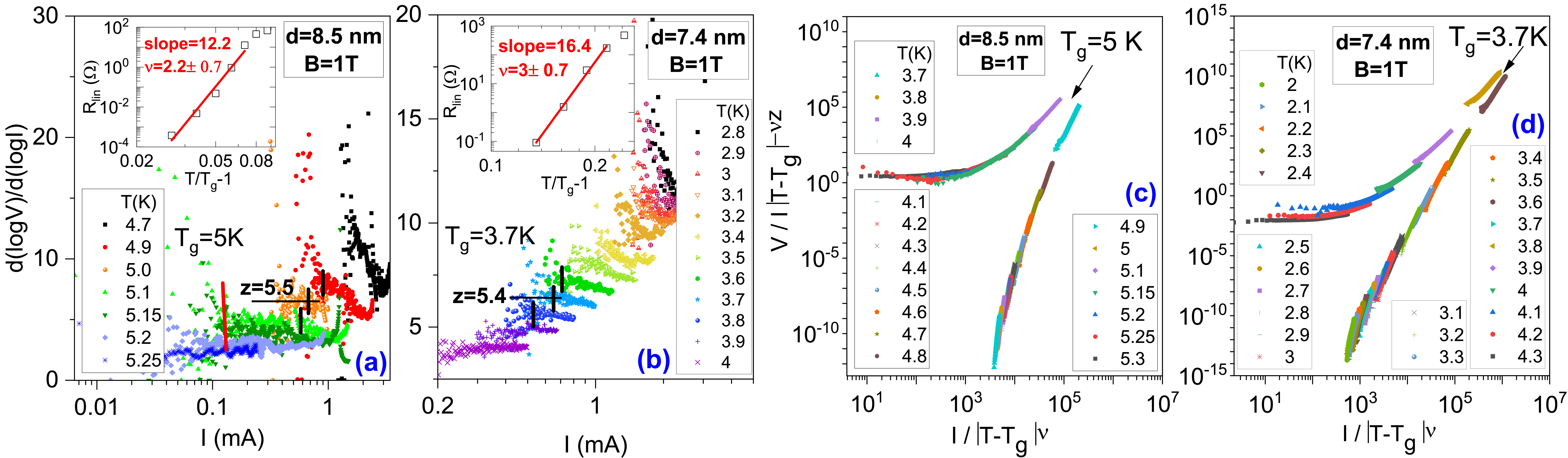}
\caption {$d(logV)/d(logI)$ versus $I$ for film with $d$ = 8.5 nm (a) and 7.4 nm (b) at the magnetic field of 1T. The black, horizontal lines indicate data sets at $T_g$, the black, vertical lines mark the onsets of creep revealed by strong pinning analysis, and the red, vertical line in (a) shows the $I_{min}$ above which finite size effect is negligible. The insets in (a) and (b) show $R_{lin}$ versus $T/T_g -1$. (c-d) Quasi-2D VG scaling of the $I-V$ curves in the indicated temperature range in the vicinity of $T_g$ demonstrating the two dynamic branches at $B$=1 T for $d$=8.5 nm (c) and 7.4 nm (d).}\label{Scaling}
\end{figure*}

The problem encountered during such analysis is the proper choice of the $T_g$ value. Let us point out first that the $T_g^R$ isotherms may be already affected by creep effects (particularly in case of thinner sample), and this invalidates scaling. Therefore, either the $T_g$ is larger than $T_g^R$, or the scaling is not valid at all in the present case.

In order to settle this issue we select the $T_g$ following the standard procedure, which involves plotting the derivative $d(\log{V})/d(\log{I})$ versus $I$, and looking for the temperature $T_g$, at which this plot follows a horizontal line, separating data sets with opposite concavities on decreasing $I$, downward at $T > T_g$, and upward at $T < T_g$ \cite{Strachan2001}. Figs. \ref{Scaling}(a) and \ref{Scaling}(b) show the results of this procedure for two thinner films at $B = 1$ T; the black horizontal lines show the selected data sets, what identifies the $T_g$ and the $z$ value, while following Eq.(\ref{Tg1}) the plot of $R_{lin}$ versus $T/T_g -1$ (on logarithmic scales, in the insets) gives value of the exponent $\nu$. The $T_g$ selected this way is substantially higher than the $T_g^R$; corresponding isotherms are marked in Fig. \ref{IVC}. Figs. \ref{Scaling}(c) and \ref{Scaling}(d) show the collapse of the data after applying scaling procedure, which appears reasonably good in first of these figures, and somewhat less satisfactory in the latter. The values of $T_g$, $z$ and $\nu$ for different magnetic fields are listed in Table \ref{table}, where we include also the data for the film with $d$=44 nm. Comparing the values of the critical exponents obtained by this procedure with these expected theoretically and reported previously for quasi-2D scaling ($z = 4 \sim 7$ and $\nu = 1 \sim 2$) \cite{Yamasaki1994,Villegas2005,Lee2010,Sun2013,Zhang2018,Song2019}, we observe that the $z$ values are within the expected range, while values of $\nu$ are consistently somewhat too high in case of two thinner samples.

However, it is important to point out here that the choice of the appropriate data sets for defining of the $T_g$ is somewhat tricky, particularly in case of the film with $d=7.4$ nm, for which the difference between upward and downward curvature is very gradual. This produces larger uncertainty of the $T_g$ value in case of the thinnest film. Moreover, the procedure may be affected by two effects. First, finite size effect may modify the shape of the isotherms, leading to Ohmic tails at current density smaller than about $J_{min} \sim k_B T / (2 \pi \Phi_0 d^2)$ \cite{Sullivan2004,Xu2009}. In the present case this translates into minimal current $I_{min} \sim 0.12$ mA for 8.5 nm film (and even smaller value for 7.4 nm film), shown in Fig.\ref{Scaling}(a) by dashed red vertical line, above which this effect is likely negligible; thus, we can safely ignore it. More problems result from the presence of creep, which affects the data at large currents as shown by green shadow in Figs.\ref{IVC}(a) and \ref{IVC}(b). In order to visualize better approximate creep boundaries at temperatures close to the $T_g$ we mark them by short, vertical, black lines in Figs. \ref{Scaling}(a) and \ref{Scaling}(b). It is clear that the presence of creep restricts the data available for scaling analysis, particularly in case of thinner film. Thus, our considerations suggest rather marginal applicability of scaling in the ultrathin Nb films, despite apparent validity suggested by Figs. \ref{Scaling}(c) and \ref{Scaling}(d).

The marginal applicability of scaling analysis does not mean that the VG phase is nonexistent in the films (vanishing resistance is clearly seen at small current values), but it indicates increasing inhomogeneity with decreasing film thickness. Various forms of disorder contribute to this effect. For example, the proximity of the niobium-silicon interface may lead to small admixture of Si ions into Nb layer, creating point defects, and the surface layer in which these defects are present plays increasingly important role with decreasing film thickness. However, we will show in the next section that the disorder which has the most impact on SC state in the films is related to imperfect fusion of polycrystalline grains. Such grains, which appear during the process of film growth, coalescence into smooth film with increasing film thickness. In very thin films this coalescence is not complete, producing amorphous regions at the grain boundaries.
This leads to increasing role of disorder with decreasing $d$, what may likely lead to fragmented superconductivity at some smaller $d$. In the present case of polycrystalline films the SC state still appears to be global at low $T$, but the disorder enhances the chance of creep appearing with increasing current.

\begin{figure}
	\centering
	\includegraphics[width=8.5cm]{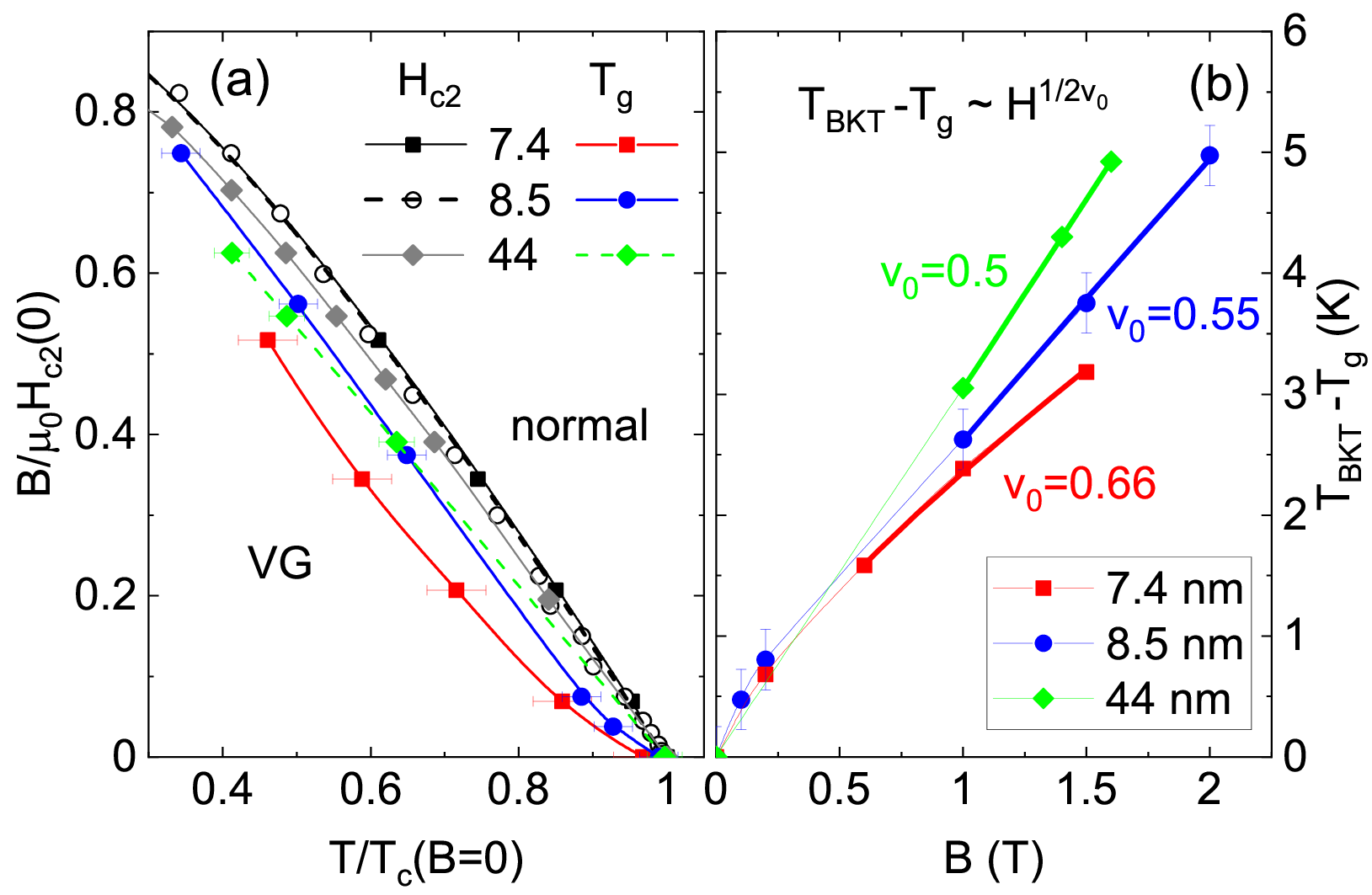}
	\caption {(a) Phase diagram for films with different $d$, with vertical and horizontal scales normalized to $\mu _0 H_{c2} (0)$ and $T_c (B=0)$, respectively. (b) $T_{BKT} -T_g$ versus magnetic field for films with different $d$; points are experimental data, lines show fits to power law with exponent $\nu_0$.}\label{PhaseD}
\end{figure}

In Fig. \ref{PhaseD}(a) we present the phase diagram for these films at small current values (below the creep boundary) with vertical and horizontal axes normalized to $\mu_0 H_{c2} (T=0)$ and $T_c (B=0)$ values, respectively; gray-black lines indicate $H_{c2}$ (determined at $R_{sq} /R_N =0.95$) and color lines indicate the $T_g$. Note that the $H_{c2}$ increases upon the decrease of $d$ (with very small difference between two thinner films), while the $T_g$ decreases, with substantial difference between two thinner films. Thus, while the disorder enhances upper critical field, it reduces markedly the range of the VG phase and expands the region of vortex liquid situated between the VG and the normal phase. It may be expected that further decrease of $d$ will result in destruction of the VG phase. This will most likely proceed via breaking of the SC state into separate islands immersed in the metallic background, leading to so-called anomalous metallic state, which we have observed in the limit of very small $d$ \cite{Zaytseva2020}.

In Fig. \ref{PhaseD}(b) we compare the dependence of the difference between $T_{BKT}$ and the $T_g$ on the magnetic field (with $T_{BKT}$ determined at zero magnetic field) for three films. It has been theoretically predicted that at small magnetic field this difference may behave as $T_{BKT} - T_g \sim H^{1/2\nu_0}$, where $\nu_0$ is the zero-field static critical exponent \cite{Fisher1991}. In the past this relation has been used to estimate static exponent value in high-temperature superconductors, and it has been argued that the values of $\nu_0$ found in these experiments, 0.66 \cite{Xu2009} and 0.5 \cite{Sullivan2010}, support the agreement with the 3D-XY model or mean-field theory, respectively. We find that in the present case the exponent evolves with the decreasing $d$, from 0.5 in thickest film, to 0.66 in thinnest one. Since our films are clearly 2D systems with substantial pinning effects, the meaning of this finding is not very clear, apart from the fact that it indicates an influence of disorder on the correlation length in the VG phase.

\begin{figure}
 	\centering
 	\includegraphics[width=7cm]{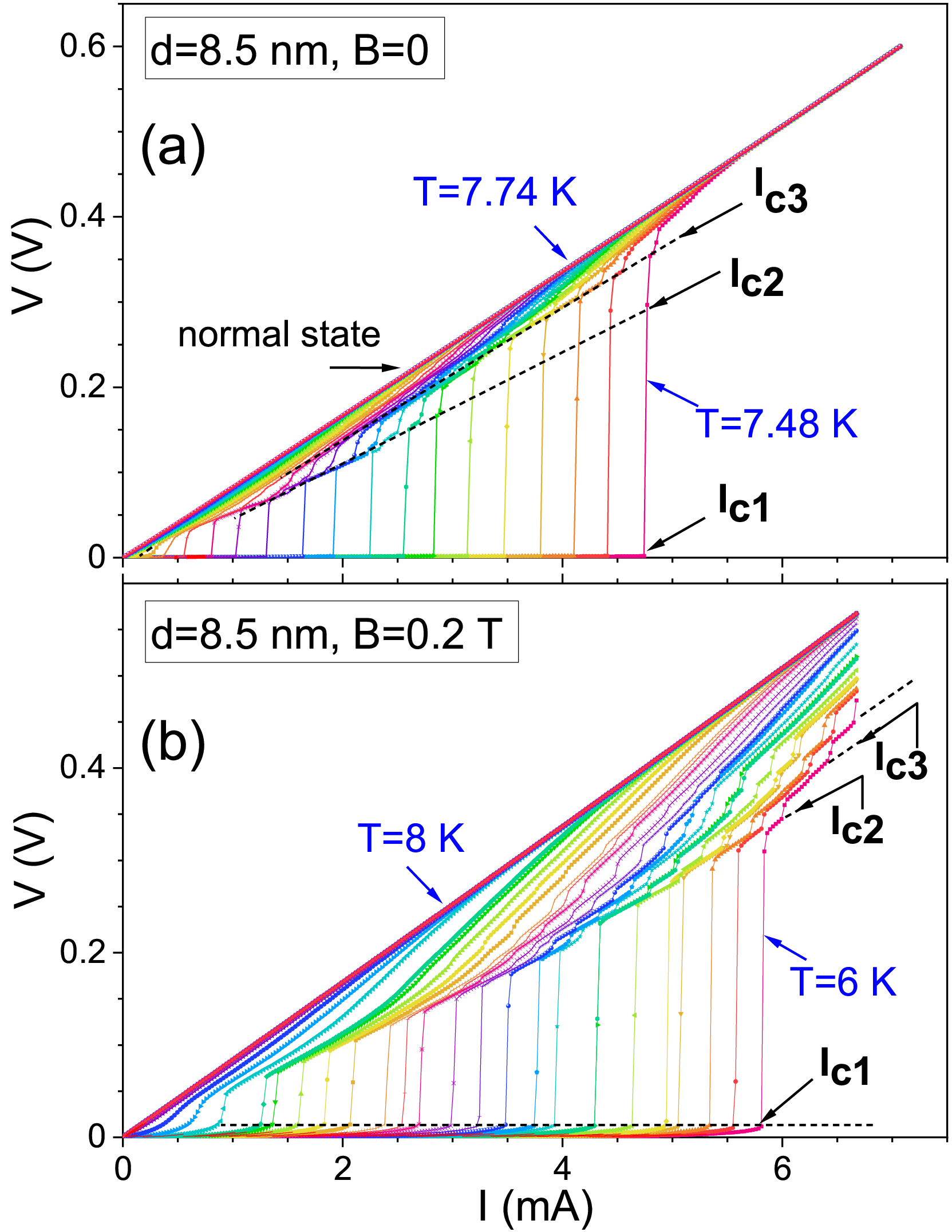}
 	\caption {Intermediate resistive states in the IV characteristics for 8.5 nm Nb film at $B = 0$ (a) and 0.2 T (b). The consecutive measurement temperatures differ by 10 mK in (a), and by 50 mK in (b). The dashed lines in the figures indicate the locations of different critical currents $(I_{cn}; n=1, 2, 3,...)$ separating successive resistive states in each $I-V$ isotherm.}\label{PLS}
 \end{figure}

Before the end of this section we return to comment briefly on the observation of intermediate resistive states, evident in Fig. \ref{IVC}(a). In Figs. \ref{PLS}(a) and \ref{PLS}(b) we display this feature in more details, at two different magnetic fields, 0 and 0.2 T, respectively. It is seen that after the current exceeds certain critical value (which we call $I_{c1}$), the resistance jumps to the normal state via several intermediate steps, each time after exceeding consecutive critical values, $I_{cn}, n=2,3,4...$. The series of jumps depend on magnetic field and temperature; in particular, small magnetic field, which broadens slightly the transition and shifts whole series of jumps to lower temperature, seems to be more suitable for observation of this effect.

This behavior is a clear manifestation of nonequilibrium properties of the vortex fluid at high current densities. In particular, the characteristic features observed here suggest occurrence of the phase slip lines (PSLs) phenomenon, usually observed at high current densities in wide, 2D superconducting films \cite{Ogushi1972,Sivakov2003,Bawa2015,Paradiso2019}. Numerical simulations suggest that the effect is caused by rivers of fast vortices and antivortices (so-called kinematic vortices) that annihilate in the middle of the sample \cite{Berdiyorov2009,Berdiyorov2014}. Nice visualization of a real effect has been obtained for films of Sn and InSn using scanning laser microscopy \cite{Sivakov2003}. The PSL effect is different from the phenomenon of phase slip centers, that occurs in narrow SC wires with width $w$ much smaller than the coherence length $\xi$ or magnetic penetration depth $\lambda$. In the present experiment we estimate $\xi(0) \approx 11$ nm and $\lambda(0)\approx 73$ nm; on increasing temperature up to $T/T_c = 0.97$ these parameters increase to about 63 nm and 430 nm, respectively, which is well below sample width ($w = 200$ $\mu$m), so that the PSL effect is a more likely cause of the observed behavior. Interestingly, while we observe this effect in film with $d = 8.5$ nm, it is absent in thinner film with similar $\xi$ and $\lambda$ values and with the exactly same lateral dimensions. This may suggest that the nature of the disorder may be an important parameter which affects the occurrence of the PSL phenomenon. We leave more detail evaluation of this effect to future studies.

\subsection {Vortex dynamics and strong pinning theory}

\begin{figure}
	\centering
	\includegraphics[width=7cm]{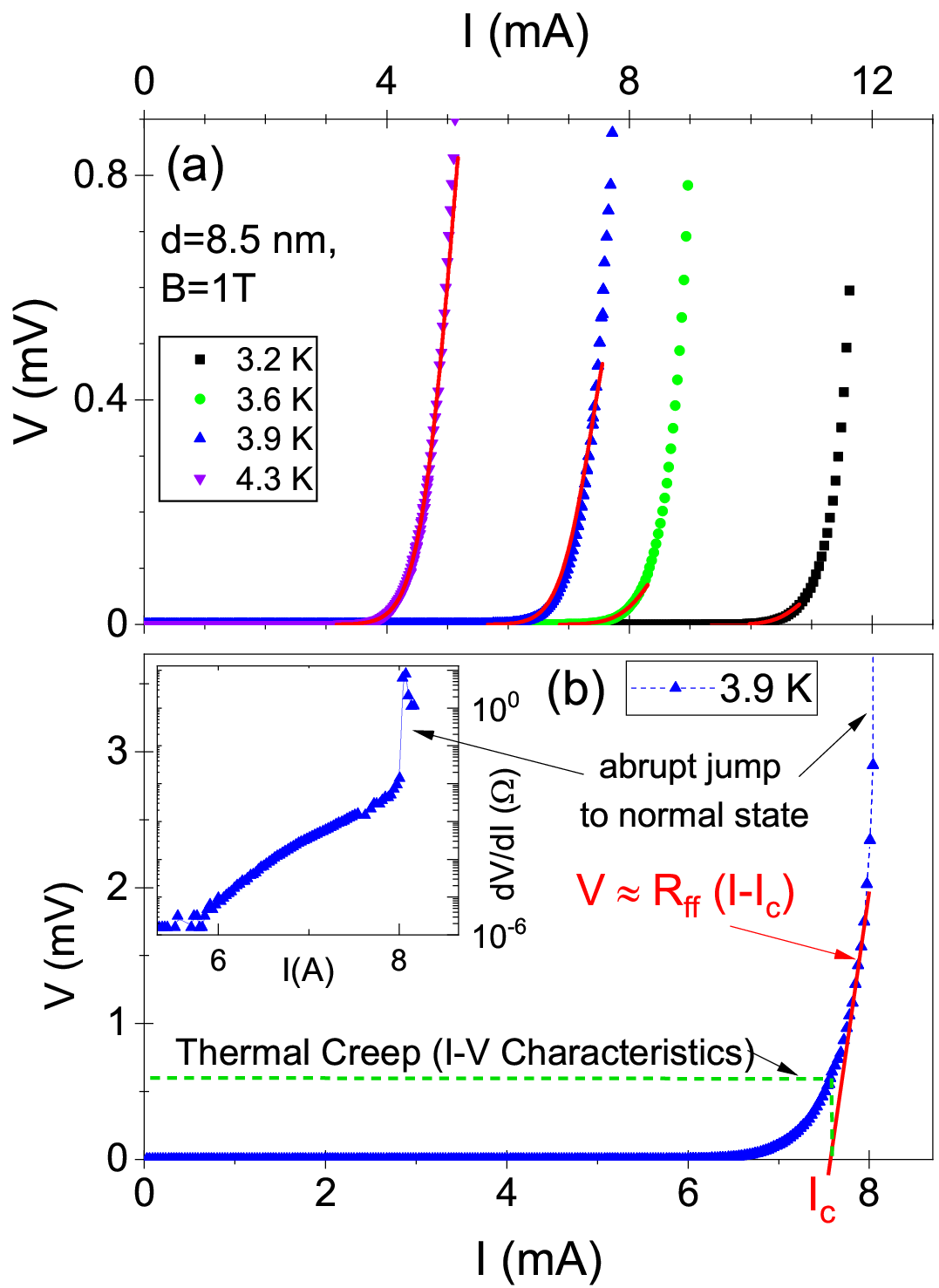}
	\caption {(a) $I-V$ characteristics measured for 8.5 nm Nb film at a fixed magnetic field $B = 1$ T and temperatures $T$ from 3.2 K to 4.3 K. Red lines are fits to the data based on the prediction of strong pinning theory \cite{Buchacek2019} (see text for details). (b) Thermal creep and excess $(I-V)$ characteristics for 3.9 K under the same conditions in (a) demonstrating the linear response at large drives preceding the abrupt jump to normal state. Thermal creep is restricted to the confined zone (dashed green rectangle) and the red, straight line defines $I_{c}$ and  $R_{FF}$. Inset in (b) shows monotonic increase of $dV/dI$ with increasing $I$.}\label{Excess}
\end{figure}

We now turn attention to a more detailed assessment of vortex dynamics in the SC state at current densities below $I_{c1}$ in the presence of external magnetic field. We focus on the regime of creep, which is marked by light green color in Fig.\ref{IVC}. Fig. \ref{Excess}(a) presents several examples of the IVC measured at different temperatures in this regime in case of film with $d = 8.5$ nm at fixed magnetic field $B = 1$T. This type of IVC, called "excess-current characteristics", have been reported for type-II SC with strongly pinned vortices \cite{Strnad1964,Berghuis1993,Xiao2002,Pace2004,Sacepe2019}. The theoretical description of strong pinning by a small density of pins at $T=0$ predicts a nearly linear-response (flux-flow) at large drives ($I > I_c$), that is shifted from $I=0$ by the critical current $I_c$ \cite{Thomann2012,Thomann2017}. At finite temperatures ($T > 0$) thermal fluctuations lead to flux creep, what modifies the observed features in three different ways: the $I_c$ is reduced to a $T$-dependent vortex depinning current, the rounding appears in the vicinity of it, and thermally assisted flux flow shows up at very small drives, $I << I_c$ \cite{Buchacek2018,Buchacek2019,Buchacek2019_2}. In Fig. \ref{Excess}(a) the red lines show the fits of the theoretical predictions according to this theory \cite{Buchacek2019_2}, which we will discuss further in the text.

The method of extraction of the $I_c$ value from experimental data is illustrated in Fig. \ref{Excess}(b). We use the linear fit to the data at large drives just before the abrupt jump to normal state in order to define the depinning drive $I_c$ and the flux-flow resistance, given by $V\approx R_{FF}(I-I_c)$, as indicated by red line (several data points just before the jump are omitted in the fit, since they may be already affected by the approach to flux flow instability). We note that in some systems with flux instability non-linear behavior of flux flow resistance has been observed, attributed to dynamic ordering of the vortex lattice in the flow region, as reported for 100 nm thick Nb films with moderate strong pinning \cite{Grimaldi2009}, or for other films \cite{Grimaldi2011}. Such feature should be manifested by the peak or broad maximum in the differential flux flow resistance preceding the flux flow instability. The inset to Fig. \ref{Excess}(b) proves that this feature is absent in our films.

\begin{figure}
 	\centering
 	\includegraphics[width=7cm]{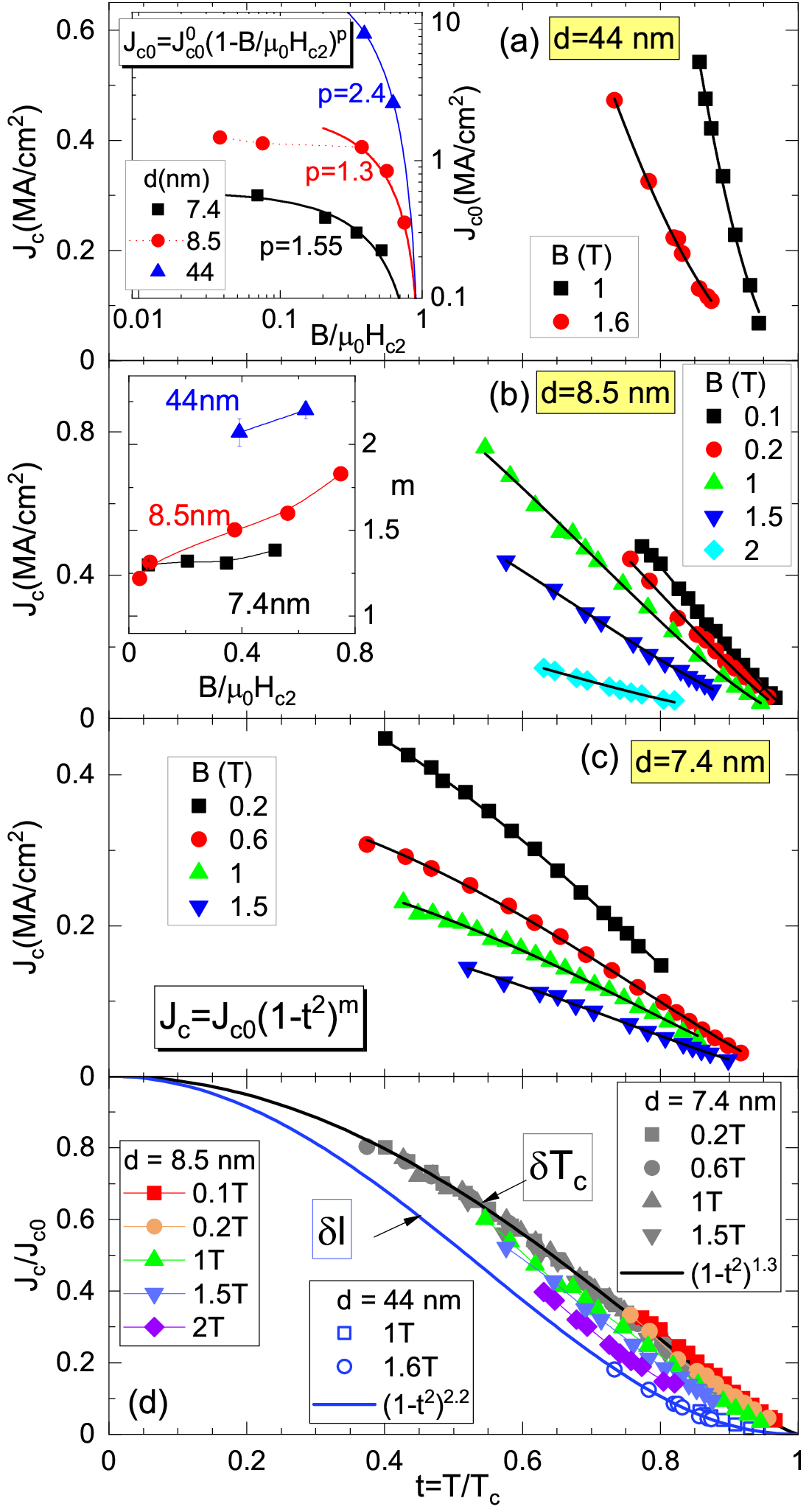}
\caption {$J_c$ versus reduced temperature $t=T/T_c$ at different magnetic fields for films with $d$ equal to 44 nm (a), 8.5 nm (b), and 7.4 nm (c). The lines show Eq. \ref{Jc} fitted to the data. (d) $J_c /J_{c0}$ versus $t$ for all films and magnetic fields. Continuous lines display Eq. \ref{Jc} with exponents $m$ equal to 2.2 (blue line) and 1.3 (black line), characteristic for $\delta l$ and $\delta T_c$ pinning, respectively.}\label{Jct}
\end{figure}

Using the $I_c$ and the films dimensions we determine the temperature dependence of the critical current density $J_c$, which is presented in Fig. \ref{Jct}(a-c) for films with different thickness measured for several magnetic fields. The data are plotted versus reduced temperature $t=T/T_c$, where $T_c$ is the SC transition temperature in the presence of given magnetic field. Black lines show that dependencies in all cases are very well described by the formula displayed in Fig. \ref{Jct}(c),

\begin{equation}
     J_c = J_{c0} (1-t^2)^m ,
     \label{Jc}
\end{equation}

where $J_{c0}$ is the $T=0$ critical current density, and $m$ is the exponent in the $t$-dependence. The behavior described by Eq.(\ref{Jc}) is frequently observed in the region of high temperatures in type II superconductors, either crystals or films \cite{Ijaduola2006,Miura2011,Maiorov2012,Sun2017,Nakamura2021}. The origins are traced to spatial variations of the coefficients in Ginzburg-Landau functional, which may be associated either with the disorder in the $T_c$ ($\delta T_c$ pinning) or with the disorder in the mean-free path of carries ($\delta l$ pinning). It has been shown that in the framework of weak collective pinning these two types of mechanisms result in different exponents $m$, 7/6 for $\delta T_c$, and 5/2 for $\delta l$ \cite{Griessen1994,note}. However, some experiments indicate that this result holds also in case of strong pinning, for example, in thin FeSeTe films \cite{Nakamura2021}.

The parameters $J_{c0}$ and $m$ extracted from the fits in the present experiment are shown in the insets to Figs. \ref{Jct}(a) and \ref{Jct}(b) as a function of the reduced magnetic field. Focusing first on the inset to Fig. \ref{Jct}(a) we observe that the value of $J_{c0}$ is high, ranging at $B = 1$ T from about 0.3 MA/cm$^2$ in the thinnest to about 8 MA/cm$^2$ in the thickest film, confirming strong, $d$-dependent pinning. The magnetic field dependence of $J_{c0}$ in case of the thinnest film is well described by a power-law dependence, $J_{c0} = J_{c0}(0) (1-B/\mu_0 H_{c2})^p$, with exponent $p = 1.55$ and the $J_{c0} (0)$ coefficient equal to about $6 \times 10^5$ A/cm$^2$. This dependence is close to mean-field theoretical relation with $p=3/2$, as recently described for other disordered SC films \cite{Sacepe2019}. Note, however, that in our experiment this is an extrapolation from relatively high temperatures ($t > 0.4$), so we cannot comment on real low-$t$ behavior. What is interesting is that in case of thicker films this relation is different, as shown in the inset to Fig. \ref{Jct}(a). In particular, the data for film with $d=8.5$ nm cannot be fitted by the same relation in the whole $H$-range. This is a first indication that the nature of pinning depends on the film thickness and magnetic field.

The exponent $m$ is seen to be dependent on the film thickness as well (inset to Fig. \ref{Jct}(b)). In the thinnest  film $m$ is close to 1.3 and it is almost $B$-independent, in the thickest film it equals to about 2.2, while in case of the film with $d = 8.5$ nm the value of $m$ is in-between, changing from the 1.25 at small $B$ up to about 2 at high magnetic field. The exponents which we obtain from the fits are close to the theoretically predicted values for $\delta T_c$ pinning in the thinnest film, and for $\delta l$ pinning in thickest film, while they suggest gradual switching from one mechanism to the other in the film with $d = 8.5$ nm. We summarize all these results by plotting in Fig.\ref{Jct}(d) the normalized critical current density, $J_c /J_{c0}$ as a function of $t$ for all films and magnetic fields. We observe that all data points for the thickest and the thinnest film are aligned nicely along two curves, $(1-t^2)^{2.2}$ and $(1-t^2)^{1.3}$, respectively, while data for film with intermediate thickness shift from $\delta T_c$ towards $\delta l$ line with increasing magnetic field.

More insight into this behavior is provided by the evaluation of the $I-V$ data using strong pinning theory introduced by Buchacek $\textit{et al.}$ \cite{Buchacek2018,Buchacek2019}, subsequently applied to interpret the experimental data in case of 2H-NbSe$_2$ single crystals and $a$-MoGe films \cite{Buchacek2019_2}. The main idea of this theory is that the thermal fluctuations modify the pinning force of vortices. At $T = 0$, in the absence of thermal fluctuations the vortex velocity $v$ is obtained from the dissipative equation of motion, in which viscous force density $-\eta v$ is given by the balance between the Lorentz force density $F_L (J, B)$ driving the vortices and pinning force density due to defects $F_{pin} (v, T)$, $-\eta v = F_L (J, B) - F_{pin} (v,T)$. Here $\eta$ is the Bardeen-Stephen viscosity per unit volume ($\eta=BH_{c2} /\rho_n c^2$, where $\rho_n$ is normal-state resistivity, and $c$ is light velocity). At small velocities $v$ and in the absence of creep $F_{pin}$ is nearly constant. This is no longer true at $T > 0$, when thermal fluctuations facilitate escape of vortices from pinning sites, so that $F_{pin}$ becomes velocity dependent in the presence of flowing current. This leads to fluctuation-modified activation energy barrier for vortex creep motion, which now depends on $F_{pin}$, $U[F_{pin}(v,T) \approx U_c [1-F_{pin}/F_c]^{\alpha}$, where $F_c$ is the critical force density, while exponent $\alpha$ and the coefficient $U_c$ depend on the type of pinning; in the strong pinning theory $\alpha=3/2$ for any smooth pinning potential. Finally, this  results in the modified current-voltage relation, which may be compared with experiment \cite{Buchacek2019_2},

\begin{equation}
V=R_{FF}(I-I_c)+V_c\left[\frac{k_B T}{U_c} ln\left(\frac{v_{th}}{v_c}\frac{V_c}{V}\right)\right]^{1/\alpha}.
\label{SPT1}
\end{equation}

\begin{figure}
	\centering
	\includegraphics[width=7.5cm]{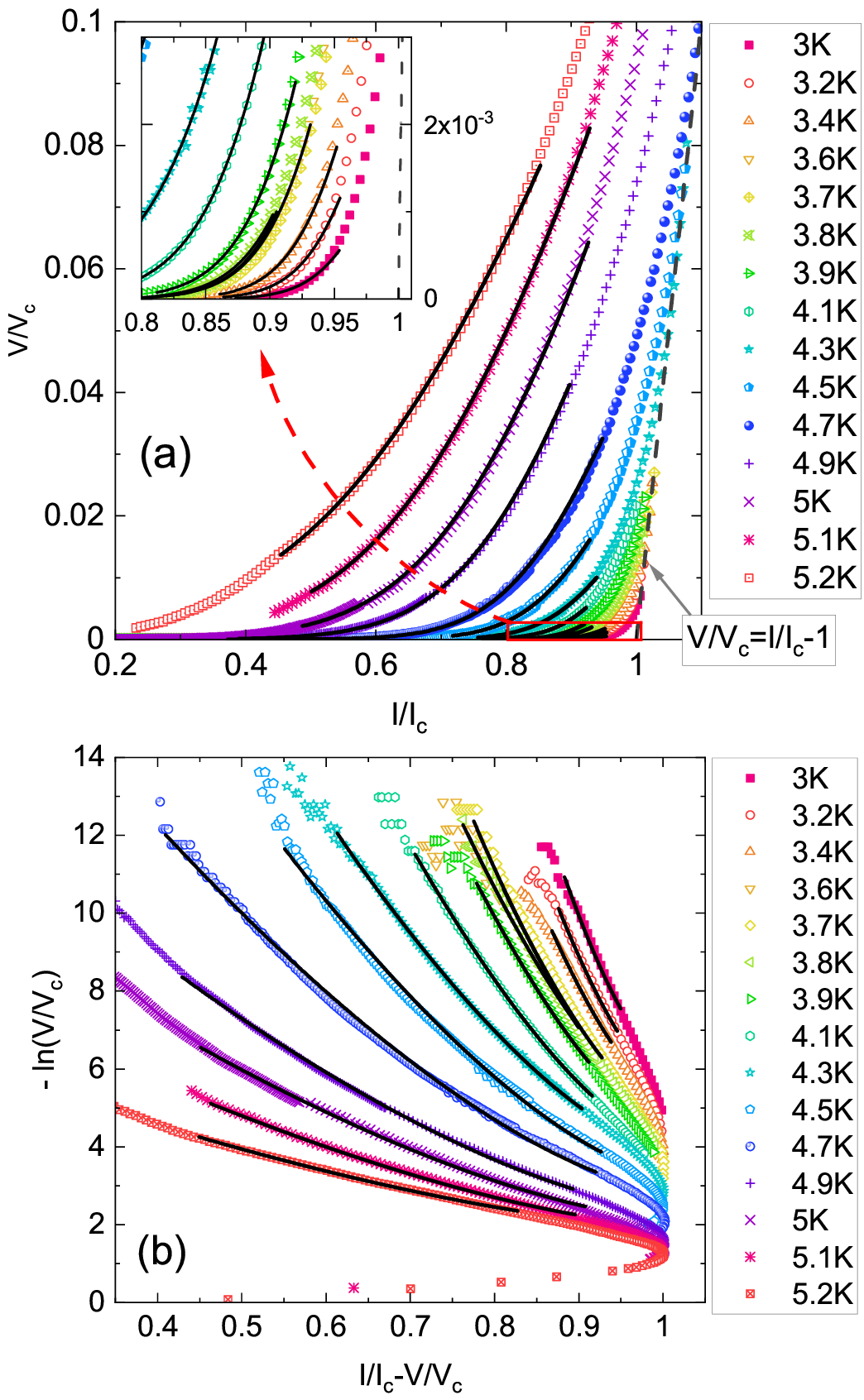}
\caption {Data for $d = 8.5$ nm film measured at $B = 1$ T at various temperatures, plotted as $V/V_c$ versus $I/I_c$ (a) and as $-ln(V/V_c)$ versus $I/I_c - V/V_c$ (b). Lines in (a) and (b) show Eqs. \ref{SPT1} and \ref{SPT2}, respectively, fitted to the data.}\label{SPTFits}
\end{figure}

Here $I_c$, $R_{FF}$ and $V_c$ are experimentally-accessible parameters (depinning current, flux flow resistance, and $V_c = R_{FF} I_c$, respectively), while $U_c /k_B T$ and $v_{th} /v_c$ are two parameters characteristic for creep behavior to be extracted from the fit of Eq. (\ref{SPT1}) to the data. The barrier $U_c$ ($T$ and $B$-dependent) should fulfil the relation $U_c \gg k_B T$, assumed during derivation of Eq. (\ref{SPT1}) ($k_B$ is the Boltzmann constant). $v_c$ is critical (free flux-flow) velocity, and $v_{th}$ is thermal velocity scale, at which vortices traverse pins fast so that the barrier slowing down the motion becomes irrelevant. The fits of Eq. (\ref{SPT1}) to the data for film with $d = 8.5$ nm at $B = 1$ T, with $\alpha$ assumed equal to 3/2, are shown in Fig. \ref{Excess}(a), and, in the rescaled coordinates $V/V_c$ versus $I/I_c$ in Fig. \ref{SPTFits}(a).

The alternative method of extraction of creep parameters is the transformation of Eq. (\ref{SPT1}) into equation

\begin{equation}
-ln\left(\frac{V}{V_c}\right) = \frac{U_c}{k_B T} \left[1-\left(\frac{I}{I_c} - \frac{V}{V_c}\right)\right]^{\alpha} + \gamma,
\label{SPT2}
\end{equation}

where $\gamma = -ln(v_{th}/v_c)$. Fig. \ref{SPTFits}(b) shows plots of this relation fitted to the same data set as in Fig. \ref{SPTFits}(a), using $\alpha = 3/2$. Both methods give consistent results for creep parameters; in all cases the extracted $U_c$ follows the condition $U_c \gg k_B T$. The fitted lines start to deviate from the data at large $I/I_c \gtrsim 0.9$, possibly indicating some admixture of flux flow to the creep \cite{exponent}.

\begin{figure}
\centering
\includegraphics[width=8.5cm]{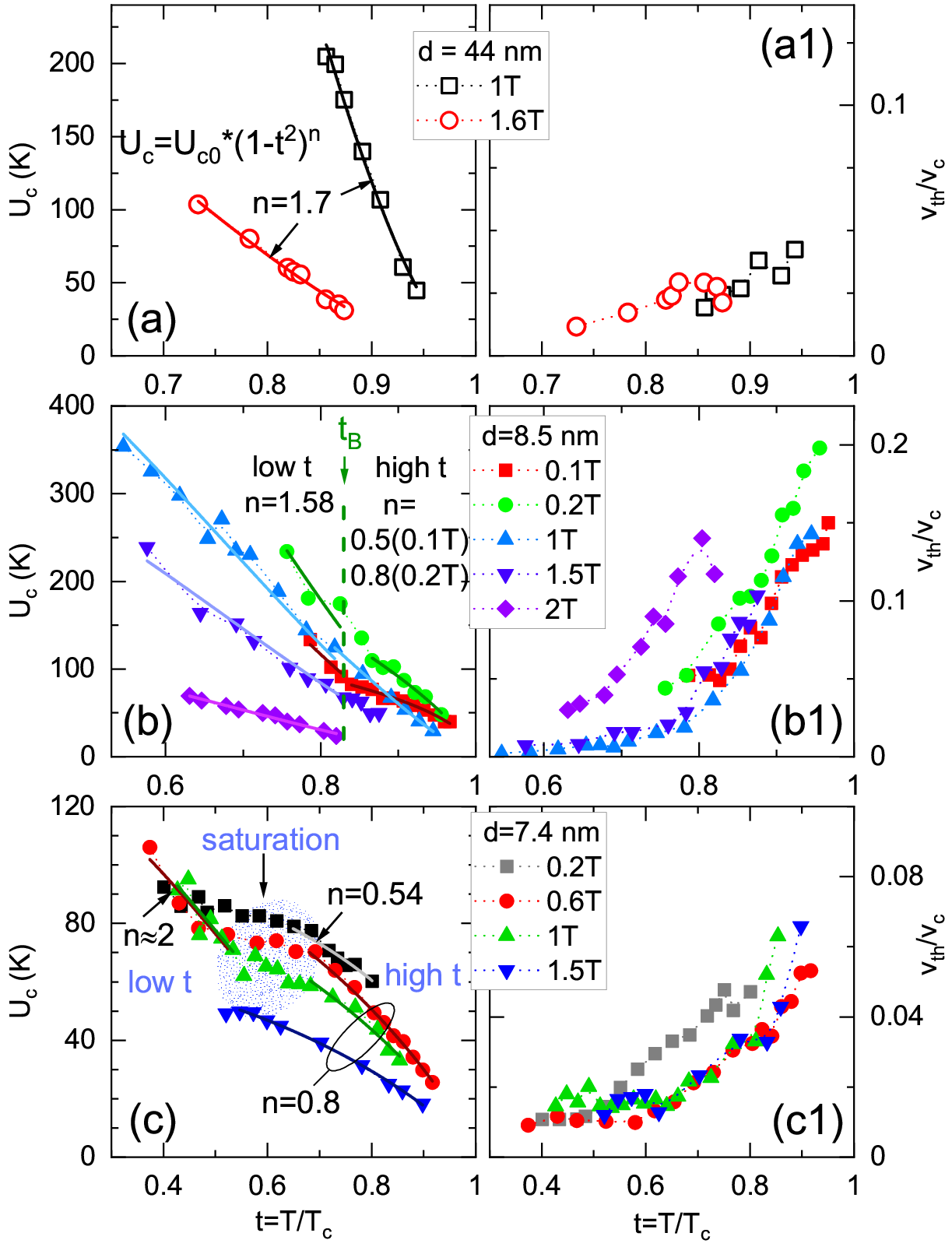}
\caption {Creep parameters $U_c$ (a-c) and $v_{th}/v_c$ (a1-c1) versus $t=T/T_c$ for $d = 44$ nm (a,a1), 8.5 nm (b,b1) and 7.4 nm (c,c1) at different magnetic fields. Continuous lines in (a-c) show fits of Eq. (\ref{Uc}), with $n$ value specified in the figure. Dashed green, vertical line in (b) shows the boundary at $t_B = 0.83$, shade in (c) shows saturation region. }\label{UcTTc}
\end{figure}

Figs. \ref{UcTTc}(a-c) and \ref{UcTTc}(a1-c1) show creep parameters, $U_c$ and $v_{th}/v_c$, respectively, extracted from the fits for different magnetic fields and plotted versus $t=T/T_c$ for samples with different $d$. The values of $v_{th}/v_c$ are rather small, but they increase with the approach to the $T_c$, in accordance with theoretical predictions \cite{Buchacek2019_2}.

The most interesting are the results for activation barriers $U_c$, which in the $T$-range studied here appear to follow functional behavior given by the relation

\begin{equation}
     U_c = U_{c0} (1-t^2)^n,
     \label{Uc}
\end{equation}

which resembles $t$-dependence of $J_c$ given by Eq. (\ref{Jc}), but with distinctly different exponents $n$. Starting from the film with $d = 44$ nm [Fig.\ref{UcTTc}(a)] we observe that $U_c (t)$ for both magnetic fields may be described by the same exponent $n = 1.7$, which is smaller than $m = 2.2$ exponent in $J_c (t)$ dependence.
In case of film with $d = 8.5$ nm [Fig.\ref{UcTTc}(b)] situation is more complicated. Unlike the $J_c (t)$ dependence, which could be fitted with a single exponent $m$ in the whole $t$-range for a given magnetic field, the $t$ dependence of the $U_c$ changes in the vicinity of $t_B=T/T_B =0.83$ (indicated by a green dashed line in the figure), from the one with exponent $n = 1.58$ for all magnetic fields at low $t$, to the one with field-dependent and much smaller $n < 1$ at hight $t$. Finally, in case of the thinnest film with $d = 7.4$ nm [Fig.\ref{UcTTc}(c)] the $U_c (t)$ dependence shows three distinct regions: high $t$ region, in which data may be fitted with Eq. (\ref{Uc}), with $n=0.54$ for the lowest magnetic field, and $n=0.8$ for all higher fields; the intermediate $t$ region, roughly for $0.5 \lesssim t \lesssim 0.7$, in which saturation of the $U_c$ is observed (marked in the figure by shadowed area), and low $t$ region, in which  the $U_c$ again increases with decreasing $t$, and $n$ is about 2 (too small amount of data points prevents more exact fitting in this range). This complicated behavior of the $U_c$ in two thinner samples is very different from much simpler form of the $J_c$. This is not surprising, because the extraction of the $U_c$ is performed in a wide range of creep, so that it provides much more information than the single value of the critical current density $J_c$.

\begin{figure}
\centering
\includegraphics[width=8cm]{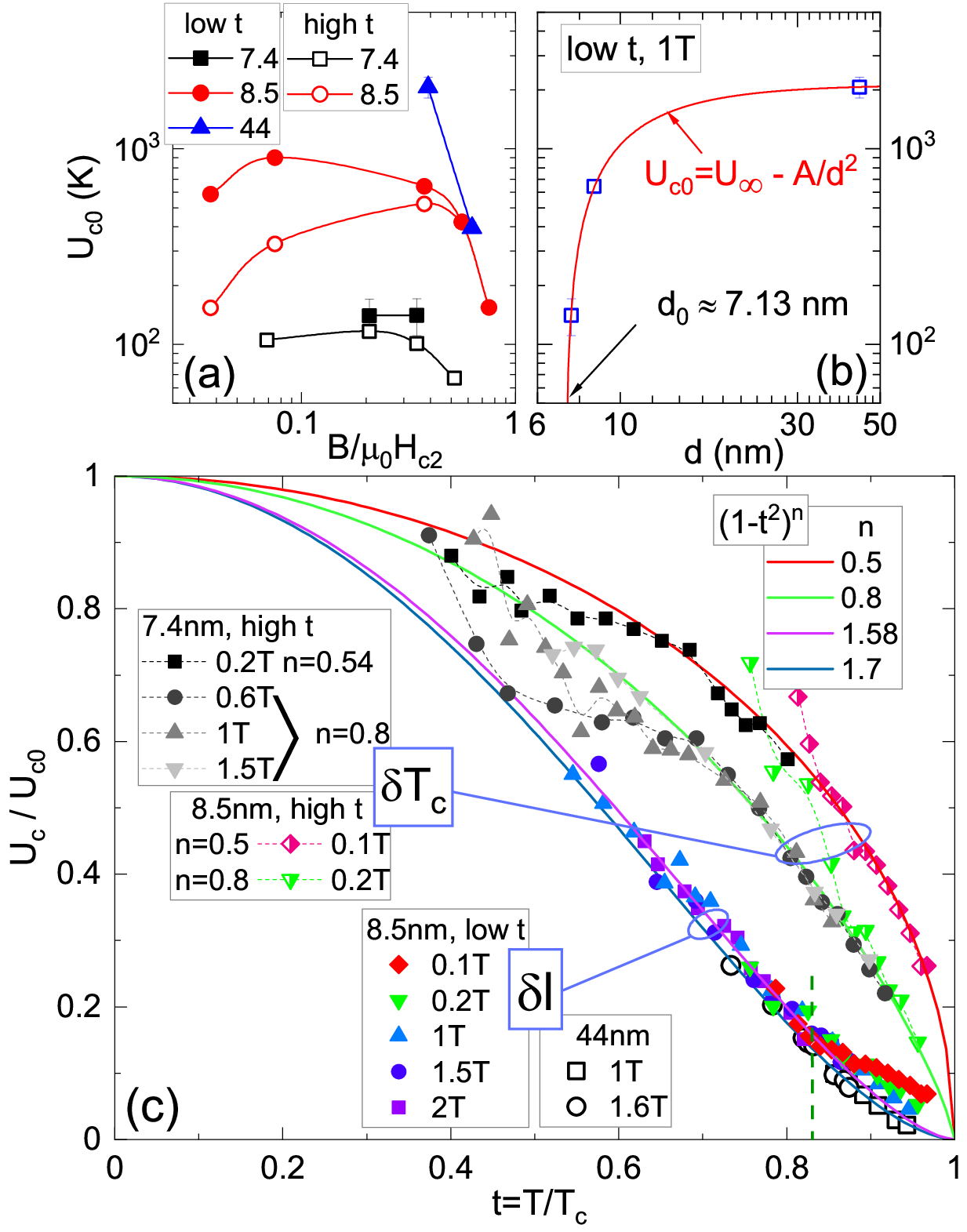}
\caption {(a) $U_{c0}$ versus $B/\mu_0 H_{c2}$ in low $t$ (solid points) and high $t$ (open points) range, for films with $d$ equal to 7.4 nm (black squares), 8.5 nm (red circles), and 44 nm (blue triangles) (unless shown, the errors are smaller than the point size). (b) $U_{c0}$ versus $d$ extracted from low-$t$ data measured at 1T; continuous line shows fit to the expression shown in the figure. (c) $U_c /U_{c0}$ versus $t$ for $d$= 44 nm (open points), 8.5 nm (low-$t$ region: solid color points, high-$t$ region: half-filled points), and 7.4 nm (grey-black solid points) for various magnetic fields. Continuous lines show function $(1-t^2)^n$, with $n=0.5$ (red), 0.8 (green), 1.58 (purple), and 1.7 (blue). }\label{UcUc0}
\end{figure}

Coefficient $U_{c0}$ ($U_c$ at $t=0$) extracted from the fits is shown in Fig. \ref{UcUc0}(a) as a function of reduced magnetic field. In case of two thinner films two different sets of $U_{c0}$ are obtained, one in high-$t$, and one in low-$t$ range (in case of thinnest film low-$t$ data could be fitted for intermediate fields only). In case of $d = 8.5$ nm film these $U_{c0}$ data sets differ by a factor of 4 at low magnetic field, but approach each other on increasing field. This immediately suggests that the mechanism of pinning at low magnetic field is different in the low-$t$ and in the high-$t$ range, while it becomes the same at high magnetic fields.

In order to get more unified picture of the observed behavior of the $U_c (t)$, in Fig. \ref{UcUc0}(c) we plot $U_c /U_{c0}$ versus $t$ for different samples and magnetic fields. The continuous lines show the function $(1-t^2)^n$ with various values of $n$, ranging from 0.5 (red line), through 0.8 (green) and 1.58 (purple), up to 1.7 (blue line). The data for film with $d = 44$ nm, shown by open points, align with blue line. All data for $d = 8.5$ nm, measured in various $B$ and fitted at low $t$ with a single function with $n=1.58$ (solid color points) align with purple line in the low-$t$ region, up to the boundary at $t_B=0.83$ (shown by dashed line). At $t > 0.83$ the data start to progressively deviate from this common line, with deviation growing with decreasing magnetic field. On the other hand, we can re-plot these high-$t$ data using $U_{c0}$ calculated from fits in high-$t$ region. This is shown by half-filled symbols for data measured at two lowest magnetic fields, 0.1 T (red diamonds), and 0.2 T (green triangles); these data nicely fit lines with $n=0.5$ (red) and $n=0.8$ (green), respectively. Finally, we add to the graph the data for film with $d = 7.4$ nm (solid black-grey points), which were fitted in the high-$t$ range: the data measured at lowest magnetic field (0.2 T) fell close to the red line, in agreement with the value of $n = 0.54$ determined by the fits, while data measured at higher fields ($n = 0.8$) align with green line. As $t$ decreases and approaches region of saturation of the $U_c$ at $t \lesssim 0.7$, the data deviate downwards from red and green lines (to avoid confusion the data at still lower $t$, with fitted $n \sim 2$, are not shown).

Following our discussion of the $J_c (t)$ behavior we propose to interpret these findings as the evidence of two distinct mechanisms of pinning, $\delta l$ in case of large $n$ ($n > 1$), and $\delta T_c$ when $n$ is small $n \thickapprox 0.5$. According to such picture, the results for thickest film indicate observation of $\delta l$ pinning, while in thinner films the pinning of $\delta T_c$-type is present in high-$t$ range, close to the $T_c$, as evidenced by red line in Fig. \ref{UcUc0}. Small increase of the exponent $n$ from 0.5 to 0.8 on increasing magnetic field in the high-$t$ range suggests that while $\delta T_c$ pinning remains the primary mechanism close to the $T_c$, some admixture of $\delta l$ pinning may occur. This is furthermore confirmed by a gradual coalescence of two different $U_{c0}$ values at high magnetic field in case of film with $d = 8.5$ nm, as shown in Fig. \ref{UcUc0}(a).

Qualitatively, such interpretation would be in line with the early theoretical descriptions of pinning mechanisms, which predict rapid increase of the $U_c /U_{c0}$ with $T$ decreasing below the $T_c$ in case of $\delta T_c$ pinning, and much slower increase in case of $\delta l$ pinning \cite{Griessen1994}. Our finding of smaller $n$ exponents for $U_c$ than $m$ exponents for the $J_c$ may also be in line with the expectation that in case of single vortex pinning the relation exists between the activation barrier and critical current density, $U_c (T,B) \sim J_c^{1/2} (T,B)$ \cite{Griessen1994}. However, the functional dependencies which we find, given by Eqs. (\ref{Jc}) and (\ref{Uc}), differ from detail predictions of Ref.\cite{Griessen1994}. This is not surprising, because these predictions assumed weak collective pinning, which is not exactly the present case. Nevertheless, more recent theoretical evaluation of several pinning models based on strong pinning theories finds qualitatively similar faster growth of $\delta T_c$ pinning than the $\delta l$ pinning with the $t$ decreasing below 1 \cite{Willa2016}, although the detail functional behavior observed in our experiment is different from predicted by this theory.

It is interesting to evaluate $d$-dependence of the $U_{c0}$ in the low-$t$ range, in which, according to our proposal, single pinning mechanism $\delta l$ exists. In Fig. \ref{UcUc0}(b) we plot $U_{c0}$ for data measured in all films at $B = 1$ T. The data indicate rapid suppression of the $U_{c0}$ with decreasing film thickness. We find that it is consistent with the dependence $U_{c0} = U_{\infty} - A / d^2$, where $U_{\infty}$ is the value of $U_{c0}$ at large film thickness and $A$ is a constant. This dependence predicts that $U_{c0} = 0$ at $d_0 \approx 7.13$ nm. While we have two few data points to verify the precise values of all constants (including exponent in $d^2$), it is clear that the $U_{c0}$ may drop to very small values at finite film thickness, so that the $\delta l$ pinning may disappear at $d$ just below the thickness of our thinnest film.

\begin{figure}
\centering
\includegraphics[width=8.5cm]{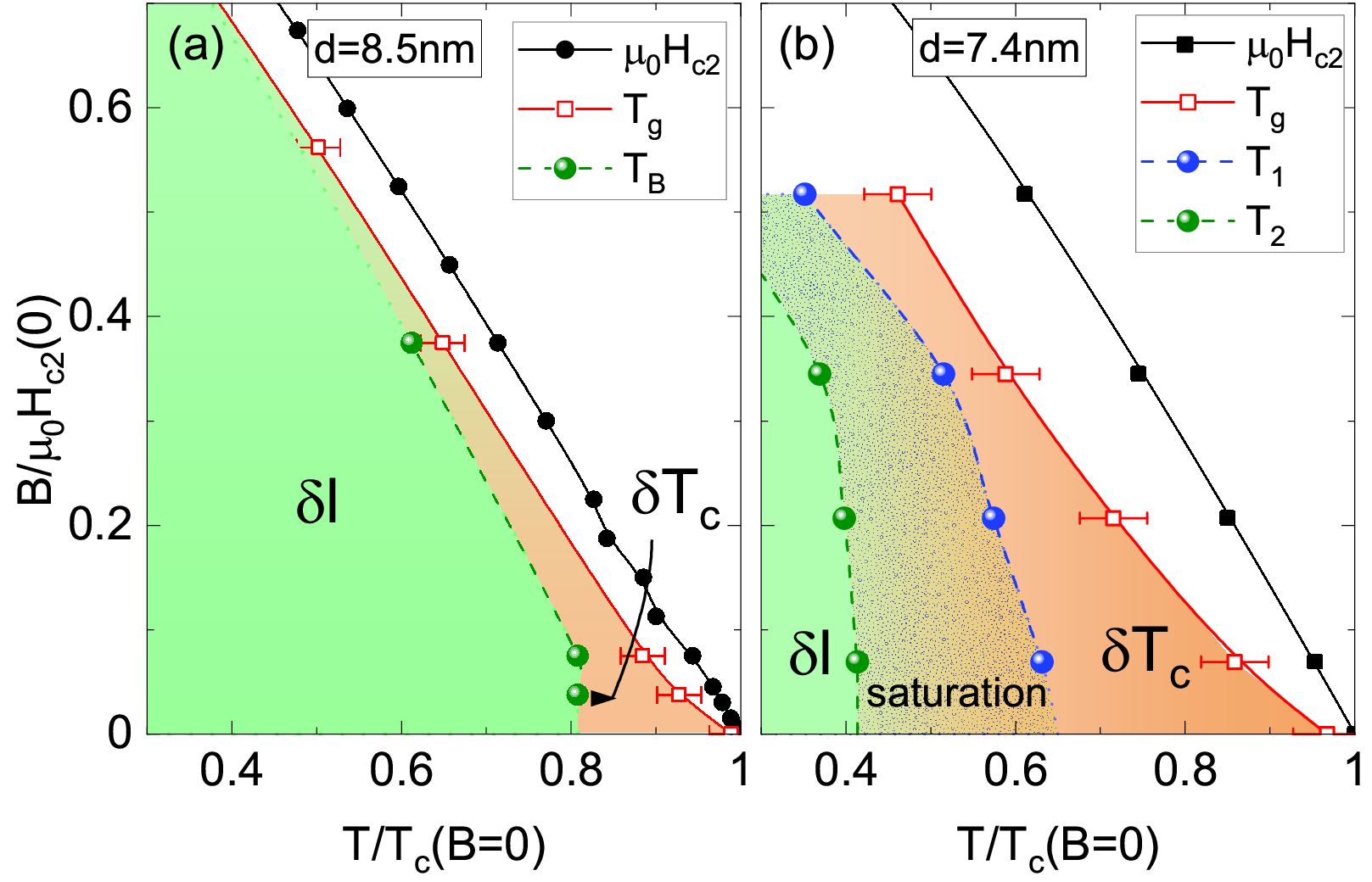}
\caption {Phase diagrams for films with $d = 8.5$ nm (a) and $d = 7.4$ nm (b), with $\mu_0 H_{c2}$ (full black points) and $T_g$ (open red points). In (a) dashed green line marks boundary at $T_B$, in (b) $T_1$ and $T_2$ mark the boundaries of the saturation region.}\label{Diag}
\end{figure}

Our observations are summarized in phase diagrams, which we propose for two thinner films in Fig.\ref{Diag} (note that the horizontal scale in these graphs is $T/T_c (B=0)$). The figure underscores dramatically different nature of pinning in two thinner films. In case of $d = 8.5$ nm film [Fig.\ref{Diag}(a)] the boundary between low-$t$ and high-$t$ regions, roughly at constant $t_B$, translates into dashed, green line marked as $T_B$, which approaches $T_g$-line on increasing magnetic field. The area in which $\delta T_c$ dominates (shown in orange color) is very narrow, and it is confined mainly to low magnetic fields and high $T$; at fields larger than 1 T it cannot be identified any more, indicating that $\delta l$ mechanism (shown by green color) becomes dominating. This feature suggests that the density of defects which cause $\delta T_c$ pinning is rather limited in this film, so that at magnetic fields $B \gtrsim 1$ T the density of vortices outweighs the density of $\delta T_c$ pins, and $\delta T_c$ mechanism stops to play a leading role. On the other hand, in case of $d = 7.4$ nm film [Fig.\ref{Diag}(b)] the $\delta T_c$ pinning persists up to the highest magnetic field measured, what suggests that the density of $\delta T_c$ pins has increased significantly. In addition, on decreasing temperature, instead of $\delta l$ pinning, we observe saturation region, delineated in the figure by two dashed lines, $T_1$ at high $T$, and $T_2$ at low $T$. Finally, at $T < T_2$ the $\delta l$ pinning most likely takes over, based on $n$ value close to 2.

We now discuss possible origins of two different pinning mechanisms, and the saturation region in thinnest film. Point defects, such as created by small interdiffusion of Si atoms into Nb layer, may be excluded from considerations, since such defects, of the atomic size, does not lead to strong pinning effects. Instead, more extended defects, of the size of about coherence length (11 nm in the present case) should be considered. In polycrystalline thin films, which grow by island growth, the most common defect structure contributing to strong pinning originates from the grain boundaries. The polycrystalline grains, of the size 3-5 nm, have been observed by us previously by transmission electron microscopy in the films deposited by magnetron sputtering, similar to the films used in present study; in fact, grain boundary scattering leads to specific dependence of the normal-state resistance on the film thickness, $R_N \sim d^{-2}$ \cite{Zaytseva2014}.

During the film deposition the island formation is initiated first, followed by coalescence of the islands into smooth film, which, however, leaves slight depression in between the grains, leading to small surface modulation. Experiments on imaging of Nb films with thickness of about 30-80 nm indicate that such modulation, of the lateral size comparable to the coherence length, and of the height not exceeding 10 \% of film thickness, forms efficient pinning sites in films \cite{Volodin2002}. Thus, we believe that the $\delta l$ pinning, which we observe in the thickest film with $d = 44$ nm, results from spatial variation of the mean free path induced by such thickness modulation. However, as the thickness of the films decreases the coalescence of the grains in the film becomes less perfect, leading to the formation of narrow amorphous regions at some grain boundaries. The size and density of such amorphous regions, randomly distributed across the film, will increase with decreasing film thickness. Such regions are a likely source of the depressed $T_c$, so they will result in $\delta T_c$ pinning mechanism.

Note that we propose to connect the origin of both pinning mechanisms to the same structural imperfections in the films, i.e. grain boundaries, however, with the decrease of film thickness some of these imperfections, randomly distributed, evolve into amorphous inclusions. This interpretation suggests that the region in the phase diagram, in which $\delta l$ pinning is the sole mechanism responsible for pinning of vortices, should be rapidly suppressed on decreasing film thickness, so that at some film thickness $d_0$ the $\delta l$ pinning will be completely replaced by a mixture of $\delta l$ and $\delta T_c$ mechanisms. This picture is consistent with the $d$-dependence of $U_{c0}$ estimated from low-$t$ region and displayed in Fig. \ref{UcUc0}(b).

Finally, we should address the origins of the saturation region in the phase diagram of the thinnest film. Since the saturation occurs close to the boundary between $\delta T_c$ and $\delta l$ regions, it is natural to suspect that the competition between two different mechanisms of pinning is responsible for this effect. In the vicinity of the boundary the activation barriers produced by these two types of pins become very close. This likely leads to frustration in the pinning landscape, what facilitates easier creep of vortices, and is reflected in saturation of the $U_c$. The effect is strong in case of thinnest film, because the density of amorphous regions at grain boundaries is large in this film. On the other hand, in thicker film with $d = 8.5$ nm much smaller density of amorphous grain boundaries produces very narrow saturation region, which is not observed in experiment.

\section{Conclusions}
	
In summary, we have used resistance and current-voltage characteristics measurements in perpendicular magnetic field to study the
vortex dynamics in thin Nb films with thickness in the range between 7.4 nm and 44 nm. Standard methods, including scaling laws, are applied to identify the VG transition temperature. While quasi-2D VG transition is confirmed in case of the thickest film, with scaling exponents firmly within theoretically expected and previously reported values, the presence of creep in thinner films complicates this identification, resulting in large uncertainty in the putative VG transition temperature and scaling exponents.

The creep regime is subsequently analyzed using recent strong pinning theory. The $T$-dependence of the activation energy for vortex pinning $U_c$, extracted from this analysis, reveals two quite distinct regimes of pinning, which we propose to identify with $\delta l$ and $\delta T_c$-type of pinning. The first of these regimes, $\delta l$, with slow increase of the $U_c /U_{c0}$ below $T/T_c =1$, is observed in thickest film. As the film thickness decreases, the second regime ($\delta T_c$) appears, characterized by faster growth of the $U_c /U_{c0}$ below $T/T_c =1$. We link these two different regimes with the grain boundaries existing in the films, which in thicker film produce carrier scattering, while in thinner films gradually evolve into amorphous inclusions, leading to local depression of the $T_c$. Experimentally obtained $T$-dependence of the $U_c$ does not strictly follow existing theoretical predictions, what calls for more theoretical studies of the subject.
	
\section*{Acknowledgments}
We are grateful to Leyi Y. Zhu and C.L. Chien (The Johns Hopkins University) for growing the films used in this study.


\begin{thebibliography}{99}

\bibitem{Fisher1991}
D. S. Fisher, M. P. A. Fisher, and D. A. Huse, Thermal fluctuations, quenched disorder, phase transitions, and transport in type-II superconductors, Phys. Rev. B {\bf 43}, 130 (1991).

\bibitem{Eley2021}
S. Eley, A. Glatz, and R. Willa, Challenges and transformative opportunities in superconductor vortex physics, J. Appl. Phys. {\bf 130}, 050901 (2021).

\bibitem{Sacepe2020}
B. Sacepe, M. Feigelman and T. M. Klapwijk, Quantum breakdown of superconductivity in low-dimensional materials, Nat. Phys. {\bf 16} 734 (2020).

\bibitem{Kapitulnik2019}
A. Kapitulnik, S. A. Kivelson, and B. Spivak, Colloquium: Anomalous metals: Failed superconductors, Rev. Mod. Phys. {\bf 91}, 011002 (2019).

\bibitem{Carbillet2020}
C. Carbillet, V. Cherkez, M. A. Skvortsov, M. V. Feigelman, F. Debontridder, L. B. Ioffe, V. S. Stolyarov, K. Ilin, M. Siegel, D. Roditchev, T. Cren, and C. Brun, Spectroscopic evidence for strong correlations between local superconducting gap and local Altshuller-Aronov density of states suppression in ultrathin NbN films, Phys. Rev. B {\bf 102}, 024504 (2020).

\bibitem{Zaytseva2020}
I. Zaytseva, A. Abaloszew, B. C. Camargo, Y. Syryanyy and M. Z. Cieplak, Upper critical field and superconductor-metal transition in ultrathin niobium films, Sci. Rep. {\bf 10}, 19062 (2020).

\bibitem{Benfatto2009}
L. Benfatto, C. Castellani, and T. Giamarchi, Broadening of the Berezinskii-Kosterlitz-Thouless superconducting transition
by inhomogeneity and finite-size effects, Phys. Rev. B {\bf 80}, 214506 (2009).

\bibitem{Venditti2019}
G. Venditti, J. Biscaras, S. Hurand, N. Bergeal, J. Lesueur, A. Dogra, R. C. Budhani, Mintu Mondal, J. Jesudasan, P. Raychaudhuri, S. Caprara, and L. Benfatto, Nonlinear I-V characteristics of two-dimensional superconductors: Berezinskii-Kosterlitz-Thouless
physics versus inhomogeneity, Phys. Rev. B {\bf 100}, 064506 (2019).

\bibitem{Weitzel2023}
A. Weitzel, L. Pfaffinger, I. Maccari, K. Kronfeldner, T. Huber, L. Fuchs, J. Mallord, S. Linzen, E. Ilichev, N. Paradiso, and C. Strunk, Sharpness of the Berezinskii-Kosterlitz-Thouless Transition in Disordered NbN Films, Phys. Rev. Lett. {\bf 131}, 186002 (2023).

\bibitem{Eley2018}
S. Eley, R. Willa, M. Miura, M. Sato, M. Leroux, M. D. Henry, and L. Civale, Accelerated vortex dynamics across the magnetic 3D-to-2D crossover in disordered superconductors, npj Quantum Mater. {\bf 3}, 37 (2018).

\bibitem{Villegas2005}
J. E. Villegas and J. L. Vicent, Vortex-glass transitions in low-Tc superconducting Nb thin films and Nb/Cu superlattices, Phys. Rev. B {\bf 71}, 144522 (2005).

\bibitem{Sun2013}
Y. Sun, J. Wang, W. Zhao, M. Tian, M. Singh, and M. H. W. Chan, Voltage-current properties of superconducting amorphous tungsten nanostrips, Sci. Rep. {\bf 3}, 2307 (2013).

\bibitem{Zhang2018}
C. Zhang, F. Hao, X. Liu, Y. Fan, T. Wang, Y. Yin and X. Li, Quasi-two-dimensional vortex-glass transition and the critical current density in TiO epitaxial thin films, Supercond. Sci. Technol. {\bf 31}, 015016 (2018).

\bibitem{Song2019}
Z. Y. Song, L. Shang, Z. Hu, J. Chu, P-P. Chen, A. Yamamoto, T-T. Kang, InN superconducting phase transition, Sci. Rep. {\bf 9}, 12309 (2019).

\bibitem{Roy2019}
I. Roy, S. Dutta, A. N. Roy Choudhury, S. Basistha, I. Maccari, S. Mandal, J. Jesudasan, V. Bagwe, C. Castellani, L. Benfatto, and P. Raychaudhuri, Melting of the Vortex Lattice through Intermediate Hexatic Fluid in an a-MoGe Thin Film, Phys. Rev. Lett. {\bf 122}, 047001 (2019).

\bibitem{Maccari2023}
I. Maccari, B. K. Pokharel, J. Terzic, S. Dutta, J. Jesudasan, P. Raychaudhuri, J. Lorenzana, C. De Michele, C. Castellani, L. Benfatto, and D. Popovi\'{c}, Transport signatures of fragile glass dynamics in the melting of the two-dimensional vortex lattice, Phys. Rev. B {\bf 107}, 014509 (2023).

\bibitem{Zaytseva2014}
I. Zaytseva, O. Abaloshev, P. D{\l}u\.{z}ewski, W. Paszkowicz, L. Y. Zhu, C. L. Chien, M. Konczykowski, and Marta Z. Cieplak, Negative Hall coefficient of ultrathin niobium in Si/Nb/Si trilayers, Phys. Rev. B {\bf 90}, 060505 (2014).

\bibitem{Blatter1994}
G. Blatter, M. V. Feigelman, V. B. Geshkenbein, A. I. Larkin, and V. M. Vinokur, Rev. Mod. Phys. {\bf 66}, 1125 (1994).

\bibitem{Ovchinnikov1991}
Yu. N. Ovchinnikov and B. I. Ivlev, Phys. Rev. B {\bf 43}, 8024 (1991).

\bibitem{vanderBeek2002}
C. J. van der Beek, M. Konczykowski, A. Abal\'{}oshev, I. Abal\'{}osheva, P. Gierlowski, S. J. Lewandowski, M. V. Indenbom,
and S. Barbanera, Phys. Rev. B {\bf 66}, 024523 (2002).

\bibitem{Thomann2012}
A. U. Thomann, V. B. Geshkenbein, and G. Blatter, Phys. Rev. Lett. {\bf 108}, 217001 (2012).

\bibitem{Thomann2017}
A. U. Thomann, V. B. Geshkenbein, and G. Blatter, Phys. Rev. B {\bf 96}, 144516 (2017).

\bibitem{Strnad1964}
A. R. Strnad, C. F. Hempstead, and Y. B. Kim, Dissipative machanism in type-II superconductors, Phys. Rev. Lett. 13, 794 (1964).

\bibitem{Kim1965}
Y. B. Kim, C. F. Hempstead, and A. R. Strnad, Flux-flow resistance in type-II superconductors, Phys. Rev. 139, A1163 (1965).

\bibitem{Berghuis1993}
P. Berghuis and P. H. Kes, Two-dimensional collective pinning and vortex-lattice melting in $a$-Nb$_{1-x}$Ge$_x$ films, Phys. Rev. B {\bf 47}, 262 (1993).

\bibitem{Xiao2002}
Z. L. Xiao, E. Y. Andrei, Y. Paltiel, E. Zeldov, P. Shuk, and M. Greenblatt, Edge and bulk transport in the mixed state of a type-II superconductor, Phys. Rev. B {\bf 65}, 094511 (2002).

\bibitem{Pace2004}
S. Pace, G. Filatrella, G. Grimaldi, and A. Nigro, Irreversible dynamics of Abrikosov vortices in type-two superconductors, Phys. Lett. A 329, 379 (2004).

\bibitem{Sacepe2019}
B. Sac\'{e}p\'{e}, J. Seidemann, F. Gay, K. Davenport, A. Rogachev, M. Ovadia, K. Michaeli, and M. V. Feigel\'{ }man, Low-temperature anomaly in disordered superconductors near $B_{c2}$ as a vortex-glass property, Nat. Phys. {\bf 15}, 48, (2019).

\bibitem{Griessen1994}
R. Griessen, Wen Hai-hu, A. J. J. van Dalen, B. Dam, J. Rector, and H. G. Schnack, Evidence for Mean Free Path Fluctuation Induced Pinning in YBa$_2$Cu$_3$O$_7$ and YBa$_2$Cu$_3$O$_8$ Films, Phys. Rev. Lett. {\bf 72}, 1910 (1994).

\bibitem{Willa2016}
R. Willa, V. B. Geshkenbein, and G. Blatter, Probing the pinning landscape in type-II superconductors via Campbell penetration depth, Phys. Rev. B {\bf 93}, 064515 (2016).

\bibitem{Buchacek2018}
M. Buchacek, R. Willa, V. B. Geshkenbein, and G. Blatter, Phys. Rev. B {\bf 98}, 094510 (2018).

\bibitem{Buchacek2019}
M. Buchacek, R. Willa, V. B. Geshkenbein, and G. Blatter, Phys. Rev. B {\bf 100}, 014501 (2019).

\bibitem{Buchacek2019_2}
M. Buchacek, Z. L. Xiao, S. Dutta, E. Y. Andrei, P. Raychaudhuri, V. B. Geshkenbein, and G. Blatter, Experimental test of strong pinning and creep in current-voltage characteristics of type-II superconductors, Phys. Rev. B {\bf 100}, 224502 (2019).

\bibitem{Demchenko2017}
I. N. Demchenko, W. Lisowski, Y. Syryanyy, Y. Melikhov, I. Zaytseva, P. Konstantynov, M. Chernyshova, and M. Z. Cieplak, Use of XPS to clarify the Hall coefficient sign variation in thin niobium layers buried in silicon, Appl. Surf. Science {\bf 399}, 32 (2017).

\bibitem{Fuchs1938}
K. Fuchs, Math. Proc. Cambridge Philos. Soc. {\bf 34}, 100 (1938).

\bibitem{Sondheimer1950}
E. H. Sondheimer, Phys. Rev. {\bf 80}, 401 (1950).

\bibitem{Mayadas1972}
A. F. Mayadas, R. B. Laibowitz, and J. J. Cuomo, J. Appl. Phys. {\bf 43}, 1287 (1972).

\bibitem{Park1986}
S. I. Park and T. H. Geballe, Phys. Rev. Lett. {\bf 57}, 901 (1986).

\bibitem{Yoshii1995}
K. Yoshii, H. Yamamoto, K. Saiki, and A. Koma, Phys. Rev. B {\bf 52}, 13570 (1995).

\bibitem{Tinkham1985}
M. Tinkham, \textit{Introduction to Superconductivity} (Krieger, Malabar, FL, 1985).

\bibitem{Hsu1992}
J. W. P. Hsu and A. Kapitulnik, Phys. Rev. B {\bf 45}, 4819 (1992).

\bibitem{Fisher1989}
M. P. A. Fisher, Vortex-Glass Superconductivity: A. Possible New Phase in Bulk High-T Oxides, Phys. Rev. Lett. {\bf 62}, 1415 (1989).

\bibitem{Altanany2023}
S.M. Altanany, I. Zajcewa, R. Minikayev and M.Z. Cieplak, Berezinskii-Kosterlitz-Thouless Transition in Ultrathin Niobium Films, Acta Phys. Pol. A {\bf 143}, 129 (2023).

\bibitem{Safar1992}
H. Safar, P. L. Gammel, D. J. Bishop, D. B. Mitzi, and A. Kapitulnik, SQUID Picovoltometry of Single Crystal Bi$_{2}$Sr$_{2}$CaCu$_{2}$O$_{8+\delta}$: Observation of the Crossover from High-Temperature Arrhenius to Low-Temperature Vortex-Glass Behavior, Phys. Rev. B {\bf 68}, 2672 (1992).

\bibitem{Wagner1995}
P. Wagner, U. Frey, F. Hillmer, and H. Adrian, Phys. Rev. B {\bf 51}, 1206 (1995).

\bibitem {Koch1989}
R. H. Koch, V. Foglietti, W. J. Gallagher, G. Koren, A. Gupta, and M. P. A. Fisher, Experimental Evidence for Vortex-Glass Superconductivity in Y-Ba-Cu-O, Phys. Rev. Lett. {\bf 63}, 1511 (1989).

\bibitem{Lee2010}
H-S. Lee, M. Bartkowiak, J. S. Kim, and H-J. Lee, Magnetic-field-induced crossover of vortex-line coupling in SmFeAsO$_{0.85}$ single crystal, Phys. Rev. B {\bf 82}, 104523 (2010).

\bibitem{Strachan2001}
D. R. Strachan, M. C. Sullivan, P. Fournier, S. P. Pai, T. Venkatesan, and C. J. Lobb, Do Superconductors Have Zero Resistance in a Magnetic Field?, Phys. Rev. Lett. {\bf 87}, 067007-1 (2001).

\bibitem{Sullivan2004}
M. C. Sullivan, D. R. Strachan, T. Frederiksen, R. A. Ott, M. Lilly, and C. J. Lobb, Zero-field superconducting phase transition obscured by finite-size effects in thick YBa$_2$Cu$_3$O$_{7-\delta}$ films, Phys. Rev. B {\bf 69}, 214524 (2004).

\bibitem{Xu2009}
H. Xu, S. Li, S. M. Anlage, C. J. Lobb, M. C. Sullivan, K. Segawa and Y. Ando, Universal critical behavior in single crystals and films of YBa$_2$Cu$_3$O$_{7-\delta}$, Phys. Rev. B {\bf 80}, 104518 (2009).

\bibitem{Yamasaki1994}
H. Yamasaki, K. Endo, S. Kosaka, M. Umeda, S. Yoshida, and K. Kajimura, Quasi-two-dimensional vortex-glass transition observed in epitaxial Bi2Sr2Ca2Cu30 thin films, Phys. Rev. B {\bf 50}, 12959 (1994).

\bibitem{Sullivan2010}
M. C. Sullivan, R. A. Isaacs, M. F. Salvaggio, J. Sousa, C. G. Stathis, and J. B. Olson, Scaling analysis of the static and dynamic critical exponents in underdoped, overdoped, and optimally doped Pr$_{2-x}$Ce$_x$CuO$_{4-y}$ films, Phys. Rev. B {\bf 81}, 134502 (2010).

\bibitem{Ogushi1972}
T. Ogushi and Y. Shibuya, Flux flow in type I and II superconducting films, J. Phys. Soc. Japan {\bf 32}, 400 (1972).

\bibitem{Sivakov2003}
A.G. Sivakov, A.M. Glukhov, A. N. Omelyanchouk, Y. Koval, P. M\"{u}ller, and A.V. Ustinov, Josephson Behavior of Phase-Slip Lines inWide Superconducting Strips, Phys. Rev. Lett. {\bf 91}, 267001 (2003).

\bibitem{Bawa2015}
A. Bawa, R. Jha, and S. Sahoo, Tailoring phase slip events through magnetic doping in superconductor-ferromagnet composite films, Sci. Rep. {\bf 5}, 13459 (2015).
		
\bibitem{Paradiso2019}
N. Paradiso, A. Nguyen, K. E. Kloss, and C. Strunk, Phase slip lines in superconducting few-layer NbSe$_2$ crystals, 2D Mater. {\bf 6}, 025039 (2019).

\bibitem{Berdiyorov2009}
G. R. Berdiyorov, A. K. Elmurodov, F. M. Peeters, and D. Y. Vodolazov, Finite-size effect on the resistive state in a mesoscopic type-II superconducting stripe, Phys. Rev. B {\bf 79}, 174506 (2009).

\bibitem{Berdiyorov2014}
G. Berdiyorov, K. Harrabi, F. Oktasendra, K. Gasmi, A. I. Mansour, J. P. Maneval, and F. M. Peeters, Dynamics of current-driven phase-slip centers in superconducting strips, Phys. Rev. B {\bf 90}, 054506 (2014).

\bibitem{Grimaldi2009}
G. Grimaldi, A. Leo, A. Nigro, S. Pace, and R. P. Huebener, Dynamic ordering and instability of the vortex lattice in Nb films exhibiting moderately strong pinning, Phys. Rev. B {\bf 80}, 144521 (2009).

\bibitem{Grimaldi2011}
G. Grimaldi, A. Leo, C. Cirillo, A. Casaburi, R. Cristiano, C. Attanasio, A. Nigro, S. Pace, and R.P. Huebener, Non-linear Flux Flow Resistance of Type-II Superconducting Films, J. Supercond. Nov. Magn. {\bf 24}, 81 (2011).

\bibitem{Ijaduola2006}
A. O. Ijaduola, J. R. Thompson, R. Feenstra, D. K. Christen, A. A. Gapud, and X. Song, Critical currents of ex situ YBa$_2$Cu$_3$O$_{7-\delta}$ thin films on rolling assisted biaxially textured substrates: Thickness, field, and temperature dependencies, Phys. Rev. B {\bf 73}, 134502 (2006).

\bibitem{Miura2011}
M. Miura, B. Maiorov, S. A. Baily, N. Haberkorn, J. O. Willis, K. Marken, T. Izumi, Y. Shiohara, and L. Civale, Mixed pinning landscape in nanoparticle-introduced YGdBa$_2$Cu$_3$O$_y$ films grown by metal organic deposition, Phys. Rev. B {\bf 83}, 184519 (2011).

\bibitem{Maiorov2012}
B. Maiorov, T. Katase, I. O. Usov, M. Weigand, L. Civale, H. Hiramatsu, and H. Hosono, Competition and cooperation of pinning by extrinsic point-like defects and intrinsic strong columnar defects in BaFe$_2$As$_2$ thin films, Phys. Rev. B {\bf 86}, 094513 (2012).

\bibitem{Sun2017}
Y. Sun, A. Park, S. Pyon, T. Tamegai, T. Kambara, and A. Ichinose, Effects of heavy-ion irradiation on FeSe, Phys. Rev. B {\bf 95}, 104514 (2017).

\bibitem{Nakamura2021}
R. Nakamura, M. Tokuda, M. Watanabe, M. Nakajima, K. Kobayashi, and Y. Niimi, Thickness-induced crossover from strong to weak collective pinning in exfoliated FeTe$_{0.6}$Se$_{0.4}$ thin films at 1 T, Phys. Rev. B {\bf 104}, 165412 (2021).

\bibitem{note}
In Ref.\cite{Griessen1994} the $J_c$ is described by somewhat more complicated equation, proportional to $(1-t^2)^m (1+t^2)^n $, with $n$ equal to -1/2 or 5/6 for $\delta l$ and $\delta T_c$ pinning, respectively (as derived based on data for YBaCuO films). We find, however, that in our case the best fits are given with exponent $n$ close to zero.

\bibitem{exponent}
In the region of large $I/I_c \gtrsim 0.9$ one approaches flux flow regime, in which one may use fits with exponent $\alpha = 1$. We have limited present discussion to $\alpha = 3/2$, because the results are complicated enough; we leave the discussion of flux flow regime to future work.

\bibitem{Volodin2002}
A. Volodin, K. Temst, C. Van Haesendonck, Y. Bruynseraede, M. I. Montero and I. K. Schuller, Magnetic-force microscopy of vortices in thin niobium films: Correlation between the vortex distribution and the thickness-dependent film morphology, Europhys. Lett. {\bf 58}, 582 (2002).
		
	\end{thebibliography}
\end{document}